\documentclass[12pt,a4paper]{article}
\pdfoutput=1
\usepackage{jheppub}
\usepackage{dsfont}
\usepackage{amssymb,amsmath}
\usepackage{epsfig}
\usepackage{epstopdf}
\usepackage{latexsym}
\usepackage{graphicx}
\usepackage{subfigure}
\usepackage{booktabs}
\usepackage{bbm}
\usepackage{cancel}
\makeatletter
\def\@fpheader{\relax}
\makeatother

\usepackage{tikz}
\usetikzlibrary{matrix,shapes.geometric,calc,backgrounds}
\tikzset{>=latex}

\def\O{{\cal O}}

\def\A{{\cal A}}
\def\l{{\ell}}

\def\T {{T_{\text{eff}}}}

\newcommand{\be}{\begin{equation}}
\newcommand{\ee}{\end{equation}}
\newcommand{\bea}{\begin{eqnarray}}
\newcommand{\eea}{\end{eqnarray}}

\subheader{\begin{flushright}
\end{flushright}}

\title{Spread of entanglement for small subsystems in holographic CFTs}
\author{Sandipan Kundu$^1$ and Juan F. Pedraza$^2$}
\affiliation{$^1$Department of Physics, Cornell University, Ithaca, NY 14853, USA}
\affiliation{$^2$Institute for Theoretical Physics, University of Amsterdam, 1090 GL Amsterdam, NL}
\emailAdd{kundu@cornell.edu}
\emailAdd{jpedraza@uva.nl}

\abstract{We develop an analytic perturbative expansion to study the propagation of entanglement entropy for small subsystems after a global quench, in the context of the AdS/CFT correspondence. Opposite to the large interval limit, in this case the evolution of the system takes place at timescales that are shorter in comparison to the local equilibration scale and thus, different physical mechanisms govern the dynamics and subsequent thermalization. In particular, we show that the heuristic picture in terms of a ``entanglement tsunami'' does not apply in this regime. We find two crucial differences. First, that the \emph{instantaneous} rate of growth of the entanglement is not constrained by causality, but rather its time average. And second, that the approach to saturation is always continuous, regardless the shape of the entangling surface. Our analytic expansion also enables us to verify some previous  numerical results, namely, that the saturation time is non-monotonic with respect to the chemical potential. All of our results are pertinent to CFTs with a classical gravity dual formulation.}

\begin{document}

\maketitle
\flushbottom

\section{Introduction}

Understanding the generation and spread of entanglement in quantum systems for generic out-of-equilibrium configurations is a topic of great interest, and currently one of the most challenging problems connecting
quantum information and statistical physics. If the system is prepared in a pure state, it will remain forever in a pure state due to unitarity. However, finite subsystems seem to thermalize
as a consequence of ergodicity.\footnote{If we consider a finite region in a system
of infinite size, the number of degrees of freedom  outside  the  region  is  much larger
than in the inside. Therefore, in a typical excited pure state the reduced density matrix for the finite
region is approximately thermal \cite{Page:1993df}.} A useful order parameter in these situations is the entanglement entropy $S_A$, which is defined as follows. We can imagine a Cauchy  surface  that  divides  the  entire  system in two subsystems, $A$ and its complement $B$, so that the total Hilbert space factorizes as $\mathcal{H}_{\text{total}}=\mathcal{H}_A\otimes\mathcal{H}_B$.\footnote{Notice that there can be multiple Cauchy surfaces resulting in the same partitioning of the Hilbert space. More concretely, this partition is specified by the (future) Cauchy horizon rather than the Cauchy surface itself \cite{Casini:2003ix}.} On the other hand, the state of the system is completely specified by its density matrix $\rho$, a self-adjoint, positive semi-definite, trace class operator. The entanglement entropy of a region $A$ with its complement $B$ is then defined as the
von Neumann entropy $S_A=-\text{tr}[\rho_A\log\rho_A]$, where $\rho_A=\text{tr}_B[\rho]$ is the \emph{reduced} density matrix of the subsystem $A$. Due to its nonlocal character, entanglement entropy could in principle reveal quantum correlations not accessible to other observables constructed from any subset of local operators $\mathcal{O}_i$.

The simplest dynamical process in which we could study the spread of entanglement is a global quench. To describe this process, we can consider the Hamiltonian (or the Lagrangian) of the system, denoted by $H_0$ (or
$\mathcal{L}_0$), and add a time-dependent perturbation of the form
\be
H_\lambda = H_0 + \lambda(t)\delta H_\Delta\qquad\rightarrow\qquad \mathcal{L}_\lambda = \mathcal{L}_0 + \lambda(t)\mathcal{O}_\Delta\,.
\ee
Here $\lambda(t)$ corresponds to an external (tunable) parameter and $H_\Delta$ (or $\mathcal{O}_\Delta$) represents a deformation of the theory by an operator of conformal dimension $\Delta$. Let us now imagine that
the perturbation is sharply peaked, i.e. $\lambda(t)\sim\delta(t)$, so that the quench is instantaneous. In this case, the process is effectively described by the injection of a uniform energy density at $t=0$ and the subsequent dynamics is dictated by the original Hamiltonian $H_0$. In a remarkable paper \cite{Calabrese:2005in}, Calabrese and Cardy showed that for $(1+1)-$dimensional CFTs as well as for some lattice models,
entanglement entropy for a large interval of length $\ell=2R$ grows linearly in time,
\be\label{linearCC}
\Delta S_A(t)=2t s_{\text{eq}}\,,\qquad t\leq R\,,
\ee
and then saturates abruptly at $t=t_{\text{sat}}=R$. Here, $\Delta S_A(t)$ denotes the difference of the entanglement entropy from that of the initial state (which is assumed to be the ground state of $H_0$), and $s_{\text{eq}}$ is the thermal entropy density of the final state. As explained in \cite{Calabrese:2005in}, these results can be easily understood in terms of causality applied to left- and right-moving EPR pairs of entangled quasiparticles emitted from the initial state. However, it is not clear if such a simple interpretation could be valid more generally, in particular, in systems with strong interactions between the pairs, which are ubiquitous in real-world many body systems.

The discovery of the AdS/CFT (or holographic) correspondence \cite{Maldacena:1997re,Gubser:1998bc,Witten:1998qj} opened the possibility to tackle the problem of entanglement propagation from a fundamental point of view. This remarkable correspondence has already been very useful in addressing problems of strongly coupled dynamics in various models, ranging from understanding aspects of Quantum Chromodynamics (QCD) to condensed matter-inspired systems \cite{CasalderreySolana:2011us,Hartnoll:2009sz}. In this context, global quenches (as the ones described above) are commonly modeled by a collapsing shell of matter in an asymptotically AdS geometry. See \cite{Danielsson:1999zt,Danielsson:1999fa,Giddings:2001ii} for early works on this topic. These gravity solutions have recently been employed to study the growth of entanglement after a global quench both in $(1+1)-$dimensional CFTs as well as in higher dimensional theories. For large subsystems, it was found that the evolution of entanglement exhibits a universal linear regime
\be\label{linearH}
\Delta S_A(t)=v_Es_{\text{eq}}A_{\Sigma}t\,,\qquad t_{\text{sat}}\gg t\gg t_{\text{loc}}\,.
\ee
In this formula, $v_E$ is interpreted as a velocity for entanglement propagation, which depends on the number of spacetime dimensions $d$ according to
\be\label{veH}
v_E=\sqrt{\frac{d}{d-2}} \left(\frac{d-2}{2(d-1)}\right)^{\frac{d-1}{d}}\leq1\,,
\ee
and $A_{\Sigma}$ is the area of the entangling region's boundary $\Sigma=\partial A$. The linear growth (\ref{linearH}) was first observed numerically in \cite{AbajoArrastia:2010yt,Albash:2010mv} and analytically in \cite{Hartman:2013qma,Liu:2013iza,Liu:2013qca}, and was later generalized to various holographic setups in \cite{Balasubramanian:2010ce,Balasubramanian:2011ur,Keranen:2011xs,Caceres:2012em,Caceres:2013dma,Hubeny:2013hz,Li:2013sia,Li:2013cja,Fischler:2013fba,Hubeny:2013dea,Alishahiha:2014cwa,Fonda:2014ula,
Pedraza:2014moa,Keranen:2014zoa,Buchel:2014gta,Caceres:2014pda,Zhang:2014cga,Keranen:2015fqa,Zhang:2015dia,Ziogas:2015aja,Dey:2015poa,Caceres:2015bkr,Camilo:2015wea,Roychowdhury:2016wca,Aref'eva:2016dmy}.
Generally speaking, $t_{\text{sat}}$ scales like the characteristic size of the region $t_{\text{sat}}\sim\ell$ while $t_{\text{loc}}$ is a local equilibration scale, which scales like the inverse of the final temperature $t_{\text{loc}}\sim1/T$. In $d=2$ one obtains $v_E=1$ as in \cite{Calabrese:2005in}, so entanglement propagates as if it were carried by a free streaming of particles moving at the speed of light. This suggests that interactions might not play a crucial role in the growth of entanglement entropy; however, recent investigations have shown that this picture fails to reproduce other holographic and CFT results, e.g. the entanglement entropy for multiple intervals \cite{Asplund:2013zba,Leichenauer:2015xra,Asplund:2015eha}. Further evidence comes from the results in higher dimensional theories. In \cite{Casini:2015zua} it was shown that in free streaming models
\be
v_E^{\text{free}}=\frac{\Gamma[\frac{d-1}{2}]}{\sqrt{\pi}\Gamma[\frac{d}{2}]}\,,
\ee
which is smaller than the holographic result (\ref{veH}) for $d\geq3$. This implies that the amount of entanglement generated in these simple models cannot account for the result in strongly coupled theories, so interactions must play a role.

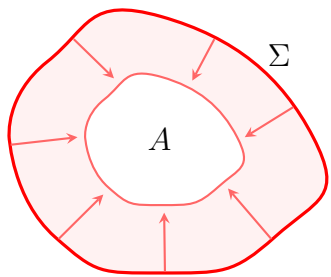
\begin{figure}
\begin{center}
\begin{tikzpicture}
\filldraw[color=red, fill=red!5, very thick] plot[smooth, tension=1] coordinates {(-2,0) (-1.4, -1.32) (0,-1.8) (1.5, -1.32) (2,0)  (0,1.6) (-1.5, 1) (-2,0)};
\filldraw[color=red!60, fill=white, thick] plot[smooth, tension=1] coordinates {(-1,0) (-0.7, -0.66) (0,-0.9)  (0.75,-0.66)  (1,0) (0,0.8) (-0.75, 0.5) (-1,0)};
\draw [color=red!60,-stealth,thick] (-1.98,-0.1) -- (-1.1,0) node [right] {};
\draw [color=red!60,-stealth,thick] (0.05,-1.78) -- (0.04,-1) node [right] {};
\draw [color=red!60,-stealth,thick] (1.74,0.4) -- (1.1,0) node [right] {};
\draw [color=red!60,-stealth,thick] (0.7,1.3) -- (0.41,0.75) node [right] {};
\draw [color=red!60,-stealth,thick] (1.45,-1.35) -- (0.88,-0.68) node [right] {};
\draw [color=red!60,-stealth,thick] (-1.35,-1.35) -- (-0.76,-0.76) node [right] {};
\draw [color=red!60,-stealth,thick] (-1.15,1.3) -- (-0.62,0.78) node [right] {};
\put(-5,-5){$A$}
\put(40,27){$\Sigma$}
\end{tikzpicture}
\caption{\small Pictorial representation of the ``entanglement tsunami'' for a subsystem $A$. The entanglement is carried by a wave that starts from the its boundary $\Sigma$ (depicted in red) and propagates inwards at a constant speed $v_E$. The shaded region has been covered by the tsunami wavefront (depicted in orange) and is now entangled with the region outside of $A$. The white region is currently not entangled but it will become at a later time.\label{tsunami}}
\end{center}
\end{figure}

Given the simplicity and universality of equation (\ref{linearH}), Liu and Suh proposed a heuristic picture for the spread of entanglement which they called ``entanglement tsunami'' \cite{Liu:2013iza,Liu:2013qca} (see Figure \ref{tsunami}). According to their interpretation, the quench generates a wave of entanglement that propagates inward from the boundary of the subsystem $A$, with the region covered by the wave becoming entangled with the outside $B$. They further conjectured that after local equilibration is achieved, $t\gtrsim 1/T$, the instantaneous rate of growth defined as
\be
\mathfrak{R}(t)\equiv\frac{1}{s_{\text{eq}}A_{\Sigma}}\frac{d S_A}{d t}
\ee
is always bounded by the tsunami velocity, i.e. $\mathfrak{R}(t)\leq v_E$. It is important to emphasize that, in spite of its name, $v_E$ is not actually a physical velocity so \emph{a priori} it is not obvious that it must be bounded by causality. More recent works have shown that for large subsystems this is indeed the case \cite{Casini:2015zua,Hartman:2015apr}. The authors of \cite{Casini:2015zua} proved it using the positivity of mutual information, while \cite{Hartman:2015apr} used inequalities of relative entropy with respect to a thermal reference state. Thus, if the conjecture on the maximum rate of growth is true, we can conclude that $\text{max}[\mathfrak{R}(t)]\leq1$.

For small subsystems, the situation is much less understood. In this case $\ell\ll1/T$ so $t_{\text{sat}}\ll t_{\text{loc}}$. The evolution of the subsystem and its thermalization take place before local equilibration is achieved and it is not clear if the growth of the entanglement should satisfy a simple law like (\ref{linearH}). Furthermore, since this linear behavior was one of the main assumptions of \cite{Casini:2015zua,Hartman:2015apr}, the bound on the maximum rate for the entanglement growth does not apply in this regime.\footnote{Another assumption of \cite{Casini:2015zua}
that is not valid for small subsystems is the fact that mutual information with the vacuum part subtracted is not generally positive definite. This can be easily checked from the analytic result of mutual information for small regions, e.g. \cite{Fischler:2012uv,Kundu:2016dyk}.} Indeed, later in this paper we will show that this is actually the case: besides the strict large interval limit, $\text{max}[\mathfrak{R}(t)]$ is not necessarily constrained by causality. We will further show that for small subsystems, the linear regime (\ref{linearH}) is absent and thus, the heuristic picture in terms of a entanglement tsunami breaks down. This is indeed expected: in this regime, the characteristic wavelength of the thermal excitations $\lambda_{\text{th}}\sim1/T$ is much larger than the size of the system, so a model of local interactions within the entangling region cannot possibly account for the growth of entanglement and its thermalization. Finally, we emphasize that our results for the growth of entanglement in the limit of small subsystems apply only for instantaneous global quenches in CFTs with holographic duals. More generally, we expect the precise growth of entanglement in this regime to be sensitive to the details of the theory and the quench itself.

This paper is organized as follows. In Section \ref{2dsec} we study the spread of entanglement for large and small intervals
based on the analytic result for holographic CFTs in $(1+1)-$dimensions. Along the way, we point out crucial differences in the corresponding behaviors and motivate a more systematic study for the propagation of entanglement for small subsystems in other holographic theories. In Section \ref{sectemp} we introduce the holographic models of global quenches that we employ in the rest of the paper: non-equilibrium states of CFTs dual to a collapsing AdS-RN-Vaidya geometries in $(d+1)-$dimensions. The motivation for studying these solutions is twofold: on one hand, it will allow us to analytically explore theories in higher dimensions, so we will be able to draw more general conclusions. On the other hand, it will give us the possibility of explaining the behavior reported in \cite{Caceres:2012em,Caceres:2014pda}, namely that for near-thermal quenches ($T\gg\mu$) the saturation time \emph{decreases} with increasing chemical potential. As mentioned in these works, understanding this peculiar behavior may be of great relevance from a phenomenological perspective, in particular for the physics of the strongly-coupled QGP. In Section \ref{secevol} we explain the approximation scheme that we use for small subregions and we perform an explicit leading-order computation for two representative boundary regions: the strip and the ball. In Section \ref{secregimes} we analyze in detail the different regimes of thermalization and we compare with the corresponding results for large subregions. We specialize to three different regimes: an initial quadratic growth, a quasi-linear growth, and the saturation. In Section \ref{secarb} we discuss some general properties of the spread of entanglement for entangling surfaces of arbitrary size, namely, the universality of the initial growth regime, and a general bound on the average velocity, $v_E^{\text{avg}}\equiv\langle\mathfrak{R}(t)\rangle$, which is obtained from bulk causality. Finally, in Section \ref{concsec} we give a brief summary of our main results and close with conclusions.

\section{Preliminaries: spread of entanglement in $(1+1)-$dimensions\label{2dsec}}

Remarkably, for holographic CFTs in $(1+1)-$dimensions the result for the evolution of entanglement entropy
after a global quench is known in a closed form \cite{Balasubramanian:2010ce,Balasubramanian:2011ur}.
This will allow us to explore, as a first example, the different regimes of the spread of entanglement for both, large and
small subsystems.

We will consider the entanglement entropy of a boundary segment of length $\ell=2R$, and introduce dimensionless variables
\be
\mathfrak{t}=2\pi Tt\,,\qquad\qquad\qquad \mathfrak{l}=2\pi TR\,,
\ee
where $T$ is the final temperature after the quench. In the final state, entanglement entropy in a $(1+1)-$dimensional CFT is given by \cite{Calabrese:2004eu}
\be
S_A=\frac{c}{3}\log\left(\frac{R}{\epsilon}\right)+\frac{c}{3}\log\left(\frac{\sinh \mathfrak{l}}{\mathfrak{l}}\right)\equiv S_{\text{vac}}+\Delta S_A\,,
\ee
where $c$ is the central charge of the theory and $\epsilon$ is a UV regulator. Notice that we have isolated two contributions: the entanglement entropy in the vacuum, $S_{\text{vac}}$, and the difference of entanglement
entropy between the thermal state and the vacuum, $\Delta S_A$. It is also useful to study the large and small interval limit of $\Delta S_A$. For $\mathfrak{l}\gg1$ we obtain
\be\label{deltasalarge}
\Delta S_A\simeq \frac{c\mathfrak{l}}{3}=s_{\text{eq}}V_A\,,
\ee
where $s_{\text{eq}}$ is the thermal entropy density,
\be
s_{\text{eq}}=\frac{\pi c T}{3}\,,
\ee
and $V_A=\ell=2R$ is the ``volume'' of the region $A$. In this limit, entanglement entropy reduces to thermal entropy and thus, satisfies the first law of thermodynamics
\be
\frac{d(\Delta E_A)}{d(\Delta S_A)}\bigg|_{\ell}=T\,,
\ee
where $\Delta E_A=\mathcal{E} V_A$ is the energy contained in region $A$, and
\be
\mathcal{E}=\frac{\pi c T^2}{6}
\ee
is the energy density of the $(1+1)-$dimensional CFT. Importantly, in this limit the entanglement entropy is an \emph{extensive} quantity since it scales with the volume of the system $V_A$. On the other hand, for small intervals, $\mathfrak{l}\ll1$, we have
\be
\Delta S_A\simeq \frac{c\mathfrak{l}^2}{18}=\frac{c\pi^2T^2\ell^2}{18}\,.
\ee
In this limit the entanglement entropy also satisfies a first law like relation for excited states \cite{Bhattacharya:2012mi,Allahbakhshi:2013rda},\footnote{Such a law is not expected to apply for generic time-dependent configurations, but it is likely to hold if the system evolves adiabatically.}
\be
\frac{d(\Delta E_A)}{d(\Delta S_A)}\bigg|_\ell=T_{\text{ent}}\,,
\ee
where, again $\Delta E_A=\mathcal{E} V_A$, and $T_{\text{ent}}$ is the so called ``entanglement temperature''. For $(1+1)-$dimensional theories $T_{\text{ent}}$ is given by
\be
T_{\text{ent}}=\frac{3}{\pi\ell}\,.
\ee
Since $T_{\text{ent}}$ is independent of the temperature, we can formally write
\be\label{defseq}
\Delta S_A=\frac{\Delta E_A}{T_{\text{ent}}}=\frac{\mathcal{E} V_A}{T_{\text{ent}}}=s_{\text{eq}}V_A\,.
\ee
Here, we have defined $s_{\text{eq}}\equiv\Delta S_A/V_A=\mathcal{E}/T_{\text{ent}}$ in analogy to (\ref{deltasalarge}). However, notice that in the limit of small subregions $s_{\text{eq}}$ is \emph{not} expected to be equal to the thermal entropy density. In particular, since $T_{\text{ent}}$ (and therefore $s_{\text{eq}}$) depend on $\ell$, the entanglement entropy is not extensive in this case.

Let us now study the time dependent setup. The evolution of the entanglement entropy after a global quench can be written as follows \cite{Balasubramanian:2010ce,Balasubramanian:2011ur}
\be
S_A(\mathfrak{t})=S_{\text{vac}}+\Delta S_A(\mathfrak{t})\,,
\ee
where $S_{\text{vac}}$ is the entanglement entropy in the vacuum and
\be\label{deltaS2d}
\Delta S_A(\mathfrak{t})=\frac{c}{3}\log\left(\frac{\sinh\mathfrak{t}}{\mathfrak{l}\,s(\mathfrak{l},\mathfrak{t})}\right)\,,
\ee
is the change in entanglement entropy following the quench. The function $s(\mathfrak{l},\mathfrak{t})$ is given implicitly by
\be\label{l2ddef}
\mathfrak{l}=\frac{\sqrt{1-s^2}}{\rho s}+\frac{1}{2}\log\left(\frac{2(1+\sqrt{1-s^2})\rho^2+2s\rho-\sqrt{1-s^2}}{2(1+\sqrt{1-s^2})\rho^2-2s\rho-\sqrt{1-s^2}}\right)\,.
\ee
with
\be
\rho\equiv\frac{1}{2}\coth \mathfrak{t}+\frac{1}{2}\sqrt{\frac{1}{\sinh^2\mathfrak{t}}+\frac{1-\sqrt{1-s^2}}{1+\sqrt{1-s^2}}}\,.
\ee
Equation (\ref{deltaS2d}) applies for any given $\mathfrak{l}$ as long as
\be
\mathfrak{t}\leq\mathfrak{t}_{\text{sat}}=\mathfrak{l}\,.
\ee
At $\mathfrak{t}=\mathfrak{t}_{\text{sat}}$ one finds that $s=1$, $\rho=\coth \mathfrak{l}$, and
\be
\Delta S_A(\mathfrak{t}_{\text{sat}})=\Delta S_{\text{eq}}=s_{\text{eq}}V_A\,.
\ee
For $\mathfrak{t}>\mathfrak{t}_{\text{sat}}$, $\Delta S_A$ remains $\Delta S_{\text{eq}}$. Unfortunately, equation (\ref{l2ddef}) cannot be inverted
analytically, so in order to extract the explicit time dependence of $\Delta S_A(\mathfrak{t})$
for $\mathfrak{t}<\mathfrak{t}_{\text{sat}}$ and fixed $\mathfrak{l}$ one must proceed numerically.
Before doing so, let us make some important remarks. For any given $\mathfrak{l}$ we can easily
compute the time-averaged entanglement velocity:
\be\label{avg2d}
v_E^{\text{avg}}=\left\langle\mathfrak{R}(t)\right\rangle=\frac{1}{s_{\text{eq}}A_\Sigma}\frac{\Delta S_A}{\Delta t}=\frac{1}{s_{\text{eq}}A_\Sigma}\frac{s_{\text{eq}}V_A}{t_{\text{sat}}}=\frac{R}{t_{\text{sat}}}=1\,.
\ee
Since the maximum growth of entanglement is bounded by its average, $\text{max}[\mathfrak{R}(t)]\geq v_E^{\text{avg}}$, one might wonder if this inequality is strictly
saturated so that $\text{max}[\mathfrak{R}(t)]=1$ for any value of $\mathfrak{l}$ or, instead, $\text{max}[\mathfrak{R}(t)]>1$ exceeding the speed of light.\footnote{We emphasize that $\mathfrak{R}(t)$ is not actually a velocity, so it is not obvious that it must obey causality.} Indeed, we will argue below that the maximum growth of entanglement generally exceeds the speed of light and it is only in the limit $\mathfrak{l}\to\infty$ that one finds $\text{max}[\mathfrak{R}(t)]\to 1$.

In order to prove this claim, it suffices to focus on the early growth regime (for a fixed value of $\mathfrak{l}$). In the limit $\mathfrak{t}\ll\mathfrak{t}_{\text{sat}}$ one finds \cite{Liu:2013qca}
\be
\rho=\frac{1}{\mathfrak{t}}+\frac{\mathfrak{t}}{12}+\cdots\,,\qquad s=\frac{\mathfrak{t}}{\mathfrak{l}}\left(\frac{1}{\mathfrak{t}}-\frac{\mathfrak{t}}{12}+\cdots\right)\,,
\ee
and
\be
\Delta S_A(\mathfrak{t})=\frac{c\mathfrak{t}^2}{12}+\O(\mathfrak{t}^4)=2\pi\mathcal{E} t^2+\cdots\,.
\ee
Therefore, at early times the instantaneous rate of growth increases linearly,
\be
\mathfrak{R}(t)=\frac{2\pi\mathcal{E} t}{s_{\text{eq}}}+\cdots\,.
\ee
Since in this regime $\mathfrak{R}(t)<1$ it is clear that the maximum rate should satisfy $\text{max}[\mathfrak{R}(t)]>1$ in order to have an average $v_E^{\text{avg}}=1$. This is true for any finite value of $\mathfrak{l}$. The strict limit $\mathfrak{l}\to\infty$ is peculiar; in this case, most part of the evolution is linear and $\mathfrak{R}(t)$ is effectively constant $\mathfrak{R}\simeq1$. We can understand this as follows: as explained in \cite{Liu:2013iza,Liu:2013qca}, one of the relevant scales that govern the regimes of thermalization is the local equilibration scale, $t_{\text{loc}}\sim1/T$. For $t\ll t_{\text{loc}}$ the growth of entanglement is quadratic but for $t\gtrsim t_{\text{loc}}$ (once the system has reached local equilibrium) the evolution is indeed approximately linear. Moreover, in $(1+1)-$dimensions this linear behavior persists all the way to the saturation time, where the entanglement equilibrates discontinuously. Altogether, the non-trivial dynamics of the system takes place over the time span $t\in[0,t_{\text{sat}}=R]$ or, equivalently, $x\equiv t/t_{\text{sat}}\in[0,1]$. In the strict limit $\mathfrak{l}\to\infty$, $t_{\text{sat}}\to\infty$ and therefore $x_{\text{loc}}\equiv t_{\text{loc}}/t_{\text{sat}}\to0$. Thus, in this limit the entire evolution is effectively linear. For small intervals $\mathfrak{l}\ll1$ and $t_{\text{sat}}\ll t_{\text{loc}}$ so a linear approximation fails.

\begin{figure}[t!]
$$
\begin{array}{cc}
  \includegraphics[angle=0,width=0.43\textwidth]{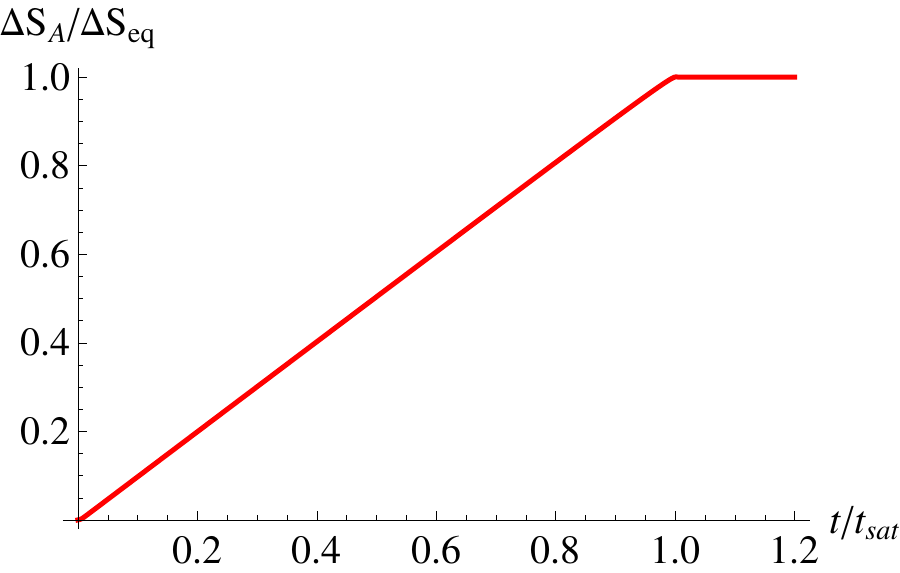} \quad &\quad \includegraphics[angle=0,width=0.43\textwidth]{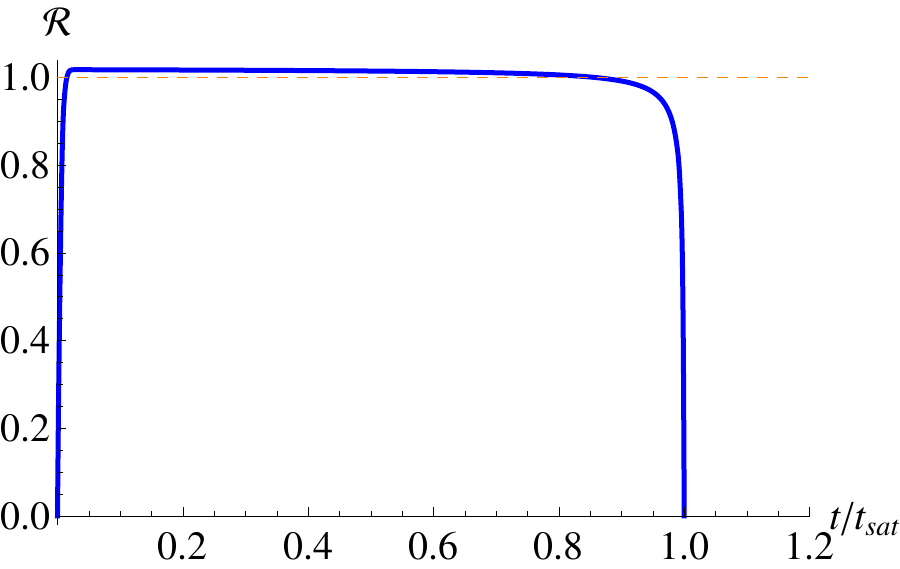}\\
  (a) \quad  & \quad (b)
\end{array}
$$
\caption{\small (a) Evolution of entanglement entropy for $TR=10^2$. For this choice of parameters $x_{\text{loc}}=t_{\text{loc}}/t_{\text{sat}}=10^{-2}\ll1$ and the growth of entanglement is approximately linear. (b) Instantaneous rate of growth for $TR=10^2$. We observe that $\mathfrak{R}(t)>1$ for $x_{\text{loc}}<x\in[0.015,0.858]$ which contradicts the conjectured bound on $\text{max}[\mathfrak{R}(t)]$. However, in the strict limit $\mathfrak{l}\to\infty$, $\mathfrak{R}(0\leq t\leq t_{\text{sat}})\to1$ (and becomes discontinuous at both $t=0$ and $t=t_{\text{sat}}$).}
\label{EE2da}
\end{figure}

\begin{figure}[ht!]
$$
\begin{array}{cc}
  \includegraphics[angle=0,width=0.43\textwidth]{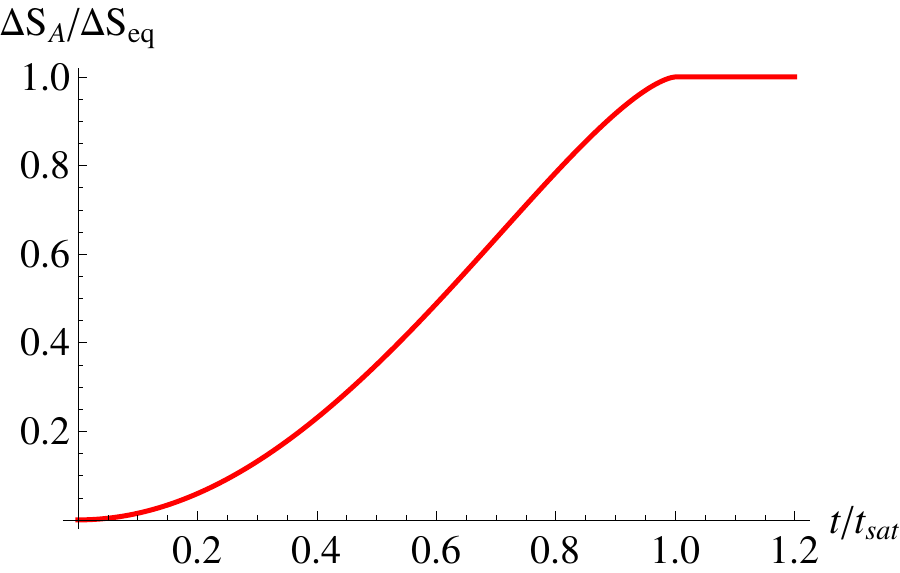} \quad &\quad \includegraphics[angle=0,width=0.43\textwidth]{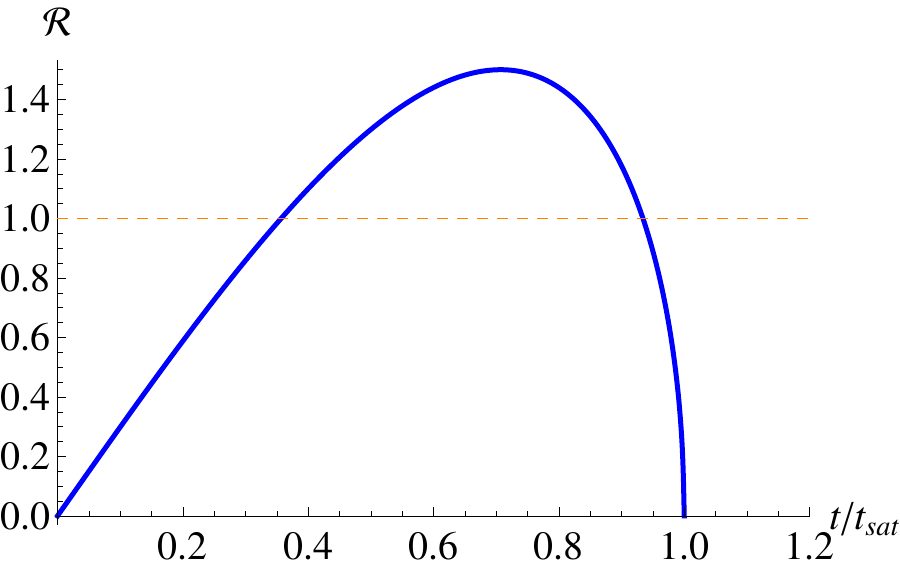}\\
  (a) \quad  & \quad (b)
\end{array}
$$
\caption{\small (a) Evolution of entanglement entropy for $TR=10^{-2}$. For this choice of parameters $x_{\text{loc}}=t_{\text{loc}}/t_{\text{sat}}=10^2>1$ and the growth of entanglement deviates from a linear behavior. (b) Instantaneous rate of growth for $TR=10^{-2}$. Our numerical results suggest a maximum rate of $\text{max}[\mathfrak{R}(t)]=3/2$.}
\label{EE2db}
\end{figure}

To add further evidence in support of these statements, we can explore numerically (\ref{l2ddef}) and study the evolution of entanglement entropy in the appropriate regimes. In Figure \ref{EE2da} we show the results for $\Delta S_A(t)$ and $\mathfrak{R}(t)$ in the large interval limit. For the plots we chose $TR=10^{2}$ so that $x_{\text{loc}}=t_{\text{loc}}/t_{\text{sat}}=10^{-2}\ll1$. As we can observe, the evolution in this case is well approximated by a straight line, and the instantaneous rate of growth $\mathfrak{R}(t)$ approaches $v_E^{\text{avg}}=1$. However $\mathfrak{R}(t)$ marginally exceeds this value for $x_{\text{loc}}<x\in[0.015,0.858]$ so the conjectured bound on $\text{max}[\mathfrak{R}(t)]$ is violated for large but finite intervals. We also observe that as we increase the size of the region, $\mathfrak{R}(t)$ becomes discontinuous both at $t=0$ and $t=t_{\text{sat}}$ in the strict limit $\mathfrak{l}\to\infty$. This agrees with the results of \cite{Liu:2013iza,Liu:2013qca} which show that, for large intervals, the approach to saturation exhibits a critical behavior akin to a first order phase transition.
In Figure \ref{EE2db} we consider the small interval limit. Here we chose $TR=10^{-2}$ so that $t_{\text{loc}}/t_{\text{sat}}=10^{2}>1$. The evolution in this case deviates from a linear behavior, which suggests that the heuristic picture in terms of a ``entanglement tsunami'' fails in this regime. The instantaneous rate of growth $\mathfrak{R}(t)$ clearly exceeds the average $v_E^{\text{avg}}$ in a good portion of the evolution: it starts off at zero, reaches a maximum $\text{max}[\mathfrak{R}(t)]>1$, and goes back to zero at $t=t_{\text{sat}}$. This indicates that the approach to saturation is generally a second order transition, rather than a first order transition, and it is only in the limit $\mathfrak{l}\to\infty$ that the discontinuous behavior manifests. Our numerical results suggest a maximum growth of $\text{max}[\mathfrak{R}(t)]=3/2$.\footnote{Regrettably, we were not able to extract this value directly from (\ref{deltaS2d})-(\ref{l2ddef}). However, we will show in Section \ref{quasilinear} that this is indeed the exact value for the maximum growth in $d=2$ dimensions.}

\section{Holographic models of global quenches in higher dimensions}\label{sectemp}

\subsection{Action and equations of motion}

Given the previous evidence, it is natural to ask if a similar behavior is also present in global quenches in higher dimensions. Here, we will consider specific models in the context of AdS$_{d+1}$/CFT$_d$ where CFT evolves from the vacuum of the theory to a state at finite temperature and/or chemical potential. The starting point is the $(d+1)-$dimensional Einstein-Hilbert action with a negative cosmological
constant coupled to a Maxwell field and an external source,
\be
S=S_0+\kappa S_{\text{ext}}\,,
\ee
where $S_0$ is given by
\begin{eqnarray} \label{action1}
S_0 = \frac{1}{8\pi G_N^{(d+1)}}\left(\frac{1}{2} \int d^{d+1} x \sqrt{-g} \left(R - 2 \Lambda \right) - \frac{1}{4} \int d^{d+1}x \sqrt{-g} F_{\mu\nu} F^{\mu\nu}
\right) \ ,
\end{eqnarray}
and $\Lambda = -\frac{d(d-1)}{2 L^2}$.\footnote{From here on we will set the AdS radius to unity $L=1$. It can be easily restored via dimensional analysis whenever necessary.} In the above $\kappa$ is a constant and $S_{\text{ext}}$ is the action of the external source, which we do not specify. This action leads to the following equations of motion
\begin{eqnarray}
&& R_{\mu\nu} - \frac{1}{2} \left(R- 2 \Lambda \right) g_{\mu\nu} - g^{\alpha\rho} F_{\rho\mu} F_{\alpha\nu} + \frac{1}{4} g_{\mu\nu}F^{\alpha\beta}F_{\alpha \beta} = 16\pi G_N^{(d+1)}\kappa T_{\mu\nu}^{\text{ext}} \ , \label{eom1} \\
&& \partial_\rho \left[ \sqrt{-g} g^{\mu\rho} g^{\nu\sigma} F_{\mu\nu} \right] = 8\pi G_N^{(d+1)}\kappa J_{\text{ext}}^\sigma\ . \label{eom2}
\end{eqnarray}
We are interested in dynamical solutions that interpolate between pure AdS and a charged AdS black hole. However, before presenting these solutions we will first study the static black hole solutions that are dual to the final state of the quench.

\subsection{Static solutions: AdS-RN\label{finalstate}}

In the absence of sources ($T_{\mu\nu}^{\text{ext}}=0$, $J_{\text{ext}}^\sigma=0$) there is a family of two-parameter black hole solutions to (\ref{eom1})-(\ref{eom2}) known as the AdS-Reissner-Nordstr\"{o}m black holes \cite{Cvetic:1999ne,Chamblin:1999tk}. For $d\ge 3$ the solutions are the following:
\begin{eqnarray} \label{RN}
&& ds^2 = \frac{1}{z^2} \left(- f(z) dt^2 + \frac{dz^2}{f(z)} + d\vec{x}^2 \right) \ , \nonumber\\
&& f(z) = 1- M z^d + \frac{(d-2)Q^2}{(d-1)} z^{2(d-1)} \ ,\\
&& A_t = Q (z_H^{d-2} - z^{d-2}) \nonumber\ ,
\end{eqnarray}
where $M$ is the mass of the black hole and $Q$ is the charge. Here, $z_H$ denotes the location of the horizon which is given by the smallest real root of
$f(z)=0$. The dual theory is a CFT that lives in $d$ spacetime dimensions and is characterized by a thermal density matrix in the grand canonical ensemble, $\rho = e^{-\beta (H -\mu q)}$, where $q$ is the total charge.
The temperature of the dual theory can be identified as the Hawking temperature of the black hole,
\be\label{tempdef}
T=-\frac{1}{4\pi}\frac{d}{dz}f(z)\bigg|_{z_H}=\frac{d}{4 \pi z_H}\left(1-\frac{(d-2)^2 Q^2 z_H^{2(d-1)}}{d(d-1)}\right) \ ,
\ee
while the chemical potential is given by
\begin{eqnarray}
\mu \equiv \lim_{z\to 0} A_t(z) = Q z_H^{d-2} \ .
\end{eqnarray}
For $d=2$ the solution takes the following form:
\begin{eqnarray} \label{RN2}
&& ds^2 = \frac{1}{z^2} \left(- f(z) dt^2 + \frac{dz^2}{f(z)} + dx^2 \right) \ , \nonumber\\
&& f(z) = 1-  M z^2 + Q^2 z^2 \log\,z \ , \\
&& A_t = Q \log\left(z_H/z\right) \ .
\end{eqnarray}
Charged solutions in $d=2$ (as the one above) have peculiar properties: the fall-off of the fields is slower than the standard case and identification of the source and the VEV are subtle \cite{Jensen:2010em} (see \cite{Perez:2015kea} for a different proposal, based on alternative boundary conditions).
To avoid these issues we will only focus on charged solutions in $d\geq3$ and consider the neutral case in $d=2$.

It is convenient to write down the metric (\ref{RN}) in the following form\footnote{Notice that (\ref{fofznew}) also includes the BTZ black hole, which is found by setting $d=2$ and $\varepsilon=1$.}
\be\label{fofznew}
f(z) = 1- \varepsilon\left(\frac{z}{z_H}\right)^d + \left(\varepsilon-1\right) \left(\frac{z}{z_H}\right)^{2(d-1)}\,,
\ee
where $z_H$ denotes the position of the horizon and $\varepsilon$ is a constant proportional to the energy density $\mathcal{E}$ \cite{Kundu:2016dyk}. In this parametrization, the temperature and chemical potential are given by
\be\label{muandT}
T=\frac{ 2(d-1)-(d-2)\varepsilon}{4\pi z_H} \ , \qquad \mu=\frac{1}{z_H}\sqrt{\frac{ (d-1)}{(d-2)}(\varepsilon-1)} \ ,
\ee
and can be inverted to obtain
\be\label{delta1}
z_H=\frac{2d}{4\pi T\left[1+\sqrt{1+\frac{d^2}{2\pi^2 a b}\left(\frac{\mu^2}{T^2}\right)}\right]}\ ,\qquad \varepsilon=a-\frac{2b}{1+\sqrt{1+\frac{d^2}{2\pi^2 a b}\left(\frac{\mu^2}{T^2}\right)}}\,.
\ee
Here, $a$ and $b$ are constants that depend only on spacetime dimensions:
\be\label{abdefs}
a=\frac{2(d-1)}{(d-2)}\ , \qquad b=\frac{d}{(d-2)}\ .
\ee
We will also define an \emph{effective temperature} $\T(T,\mu)$, which will play a crucial role:
\begin{align}\label{Teff1}
\T\equiv \frac{d}{4\pi z_H}=\frac{T}{2}\left[1+\sqrt{1+\frac{d^2}{2\pi^2 a b}\left(\frac{\mu^2}{T^2}\right)}\right]\ .
\end{align}
From the definition it follows that $\T$ interpolates between $\T\propto T$ and $\T\propto\mu$ as one goes from $\mu/T\ll1$ to $\mu/T\gg1$, so it effectively serves as a measure of the dominant scale in the theory. Specifically, for $\mu/T\ll1$ we have that
\be\label{zhthermal}
\T=T\left[1+\frac{d^2}{8\pi^2ab}\left(\frac{\mu^2}{T^2}\right)+\mathcal{O}\left(\frac{\mu^4}{T^4}\right)\right]\,.
\ee
In the opposite limit we find
\be\label{zhquantum}
\T=\frac{\mu d (d-2)}{2\pi\sqrt{2d(d-1)}}\left[1+\frac{2\pi }{d-2} \sqrt{\frac{a}{2b}}\left(\frac{T}{\mu
}\right)+\mathcal{O}\left(\frac{T^2}{\mu^2}\right)\right]\,.
\ee

Finally, we can express the various thermodynamic quantities solely in terms of $\T$ and $\varepsilon$. For instance, the temperature and chemical potential can be now written as
\be
T=\left(\frac{ 2(d-1)-(d-2)\varepsilon}{d}\right)\T\,,\qquad \mu=\sqrt{\frac{(d-1)}{(d-2)}(\varepsilon-1)}\left(\frac{4 \pi  \T}{d}\right)\,.
\ee
Similarly, the energy, entropy and charge densities are given by
\be\label{energydens}
\mathcal{E}=\frac{(d-1)\varepsilon}{16\pi G^{(d+1)}_{N}}\left(\frac{4 \pi  \T}{d}\right)^d\,,
\ee
\be
s=\frac{1}{4G^{(d+1)}_{N}}\left(\frac{4 \pi  \T}{d}\right)^{d-1}\,,
\ee
and
\be
\rho=\frac{(d-2)}{8 \pi  G^{(d+1)}_{N}}\sqrt{\frac{(d-1)}{(d-2)}(\varepsilon-1)}\left(\frac{4 \pi  \T}{d}\right)^{d-1}\,,
\ee
respectively. Together, they satisfy the first law of thermodynamics $d\mathcal{E}=Tds+\mu d\rho$.

\subsection{Collapsing solutions: AdS-RN-Vaidya}

Time-dependent solutions to (\ref{eom1})-(\ref{eom2}) describing the formation of a charged black hole have been studied in a number of works, e.g. \cite{Caceres:2012em,Caceres:2013dma}. The metric in this case is given by the AdS-RN-Vaidya solution\footnote{The AdS-RN-Vaidya solution in $d=2$ have the same issues as the static AdS-RN, hence we will only consider charged solutions in $d\geq3$. The form of (\ref{fzv}) is valid in $d=2$ provided that $q(v)=0$.}
\bea\label{vaid1}
&& ds^2=\frac{1}{z^2} \left( - f(z,v) dv^2 - 2 dv dz + d\vec{x}^2 \right) \ ,\\
&& f(z,v) = 1 - m(v) z^{d} + \frac{(d-2)q(v)^2}{(d-1)} z^{2(d-1)}\,,\qquad\text{for}\quad d\geq3\,,\label{fzv}
\eea
and is sourced by a $(d+1)-$dimensional infalling shell of charged null dust, $T_{\mu\nu}^{\text{ext}}\sim k_\mu k_\nu$ with $k^2=0$.
The explicit form of the vector field $A_\mu(v)$ will not play any role in our
discussion, so we will not transcribe it here. The metric (\ref{vaid1}) is written in terms of Eddington-Finkelstein coordinates, so that $v$ labels ingoing null trajectories. This variable is
related to the standard $t$-coordinate through
\be\label{efcoords}
dv=dt-\frac{dz}{f(z,v)}\,.
\ee
The mass $m(v)$ and
charge $q(v)$ are two functions that capture the information of the black hole formation. On physical grounds, $m(v)$  and $q(v)$ should interpolate between zero
in the limit $v\to -\infty$ (corresponding to pure AdS) and a constant value in the limit $v\to \infty$ (corresponding to an RN-AdS black hole). The final values
should not give rise to a naked singularity but, other than that, the mass and charge functions are in principle arbitrary.\footnote{However, there are stronger
constrains on $m(v)$ and $q(v)$ if we want to respect strong subadditivity in the boundary theory \cite{Caceres:2013dma}.}

One might wonder whether such a solution could be obtained from an actual collapse in asymptotically AdS space, i.e. for a specific source $S_{\text{ext}}$. Indeed, interesting steps in this direction were given in \cite{Bhattacharyya:2009uu}. In this paper, the authors studied a collapse of a massless scalar field in the so-called ``weak field expansion''. For fast quenches, and at the leading order in the perturbation, the solutions they found take the form of a Vaidya geometry (\ref{vaid1}), with a particular form of the metric that depends on the scalar profile. In the dual field theory, this corresponds to a global quench by a marginal operator, where the corresponding coupling is the small parameter in which the perturbation is carried out. Thus, at least in this approximation, the results of \cite{Bhattacharyya:2009uu} validate the phenomenological studies based on Vaidya backgrounds from a first principle computation. This approach was employed in \cite{Caceres:2014pda}, to the case of scalar collapse coupled to a Maxwell field.\footnote{It is also worth emphasizing that thin-shell limit of the Vaidya solution is in perfect agreement with numerical simulations of scalar collapse \cite{Garfinkle:2011hm,Garfinkle:2011tc}.}

Before proceeding further, let us parametrize the solution in a slightly different way. Instead of using the functions $m(v)$ and $q(v)$ we will rewrite $f(z,v)$
in terms of the apparent horizon $z_H(v)$ and an auxiliary function $\varepsilon(v)$ according to
\be
f(z,v) = 1-\varepsilon(v)\left(\frac{z}{z_H(v)}\right)^d + \left(\varepsilon(v)-1\right) \left(\frac{z}{z_H(v)}\right)^{2(d-1)}\,.
\ee
This expression is the equivalent of (\ref{fofznew}) now in the time dependent scenario,
assuming that we upgrade $T\to T(v)$ and $\mu\to \mu(v)$. Here we are defining the function $T(v)$ as
\be
T(v)\equiv-\frac{1}{4\pi}\frac{d}{dz}f(z,v)\bigg|_{z_H(v)}=\frac{2(d-1)-(d-2)\varepsilon(v)}{4\pi z_H(v)} \ .
\ee
However, strictly speaking the function $T(v)$ can only be identified with the physical temperature in the limits $v\to-\infty$ and $v\to\infty$, which
correspond to the initial and final states, respectively. Away from this two limits the system is out-of-equilibrium and the thermodynamics is not well defined.
Similarly, the function $\mu(v)$ is defined as
\be
\mu(v)\equiv\frac{1}{z_H(v)}\sqrt{\frac{(d-1)}{(d-2)}(\varepsilon(v)-1)} \ .
\ee
We can identify two special cases:
\begin{enumerate}
\item Thermal quench: in this case $\mu(v)=0$ which means $\varepsilon(v)=1$.
\item Extremal quench: in this case $T(v)=0$, which implies $\varepsilon(v)=\frac{2(d-1)}{d-2}$.\footnote{This case is often referred to as an electromagnetic quench \cite{Albash:2010mv}. For $d=3$, due to the electric-magnetic duality, this is equivalent to turn on a magnetic field in the dual CFT.}
\end{enumerate}
It will also prove useful to define the function
\begin{align}
\T(v) \equiv \frac{d}{4\pi z_H(v)}=\frac{T(v)}{2}\left[1+\sqrt{1+\frac{d^2}{2\pi^2 a b}\left(\frac{\mu(v)^2}{T(v)^2}\right)}\right]\ ,
\end{align}
which interpolates between the initial and the final effective temperature (\ref{Teff1}).

\subsubsection{Instantaneous quenches: thin shell limit\label{secquenches}}

We will work in the limit where the mass and charge functions change instantaneously: $m(v)=M\,\theta(v)$ and $q(v)=Q\,\theta(v)$, respectively.
This can be achieved by considering an infalling shell of null dust with infinitesimal thickness, which is referred to as the thin shell limit. Naively, one
might think that a thin shell would lead to an instantaneous thermalization of the field theory observables, since in this case $T(v)=T\,\theta(v)$ and
$\mu(v)=\mu\,\theta(v)$. This statement is true for one-point functions of local operators, e.g. one finds that $\langle T_{\mu\nu}(t)\rangle\sim \langle T_{\mu\nu}^{final}\rangle\,\theta(t)$. On the other hand, non-local
observables such as two-point functions and entanglement entropies actually take finite time before reaching equilibrium so they provide a more complete
information of the thermalization process.

In the thin shell limit the function $f(z,v)$ acquire the general form
\be\label{fstep}
f(z,v) = 1 - \theta(v) g(z) \ , \qquad g(z)= \varepsilon\left(\frac{z}{z_H}\right)^d - \left(\varepsilon-1\right) \left(\frac{z}{z_H}\right)^{2(d-1)} \ ,
\ee
where $z_H$ and $\varepsilon$ are related to the final temperature and chemical potential according to (\ref{delta1}). It will be useful to expand (\ref{fstep}) and define the following two kind of quenches:
\begin{enumerate}
\item Near-thermal quenches ($T\gg\mu$):
\be\label{gnthermal}
g(z)=  \left(1+\frac{(d-2) d^2\mu^2}{16\pi^2T^2}\right)\left(\frac{4\pi T z}{d}\right)^d-\frac{(d-2)d^2\mu^2}{16\pi ^2(d-1)T^2}\left(\frac{4\pi T
z}{d}\right)^{2(d-1)}+\mathcal{O}\left(\frac{\mu^4}{T^4}\right)\,.
\ee
\item Near-extremal quenches ($T\ll \mu$):
\be\label{gnextremal}
\begin{split}
&g(z)= \frac{2 (d-2)^{d-1}}{d^{d/2}(d-1)^{d/2-1}}\left(1+\frac{2 \pi  d^{1/2} T }{(d-1)^{1/2}\mu}\right)(\mu  z)^d \\
&\qquad\qquad-\frac{(d-2)^{2 d-3}}{d^{d-2}(d-1)^{d-1}}\left(1+\frac{4\pi (d-1)^{1/2}T}{d^{1/2}\mu }\right)(\mu
z)^{2(d-1)}+\mathcal{O}\left(\frac{T^2}{\mu^2}\right)\,.
\end{split}
\ee
\end{enumerate}
In both cases we have only kept the leading order corrections to the thermal and extremal quenches, respectively. Physically, the main difference between these two processes is the nature of the relevant excitations: in the first case the evolution of the system is dominated by thermal fluctuations, while in the second case it is driven by quantum fluctuations.

\section{Evolution of entanglement entropy\label{secevol}}

\subsection{General considerations for AdS-RN-Vaidya}

We are interested in computing entanglement entropy in the boundary CFT. In the context of the AdS/CFT correspondence, entanglement entropy of a region $A$ is computed by means of the Ryu-Takayanagi prescription \cite{Ryu:2006bv}, according to which:
\be \label{rt}
S_A = \frac{1}{4 G_N^{(d+1)}} {\rm min} \left[ {\rm Area} \left(\Gamma_A \right)\right] \ ,
\ee
where $G_N^{(d+1)}$ is the bulk Newton's constant and $\Gamma_A$ is a $(d-1)$-dimensional surface in the bulk such that $\partial \Gamma_A = \partial A = \Sigma$. This proposal has been
generalized to time dependent backgrounds in \cite{Hubeny:2007xt}. In this case,
\be \label{hrt}
S_A = \frac{1}{4 G_N^{(d+1)}} {\rm ext} \left[ {\rm Area} \left(\Gamma_A \right)\right] \ ,
\ee
where the condition for minimal surfaces is now replaced by extremal surfaces.

We will compute the entanglement entropy for two representative boundary regions:
\begin{itemize}
  \item A $(d-1)-$dimensional strip of width $\ell$, specified by
  \begin{equation}
x\equiv x_1 \in \left[-\frac{\ell}{2},\frac{\ell}{2}\right]\,,\qquad  x_i\in \left[-\frac{\ell_\perp}{2},\frac{\ell_\perp}{2}\right],\quad i=2,...,d-2
\end{equation}
with $\ell_\perp \rightarrow \infty$. The corresponding extremal surface $\Gamma_A$ is invariant under translations in the transverse directions, $\vec{x}_\perp$. Therefore, without loss of generality,
we can parameterize it with two functions, $x(z)$ and $v(z)$, satisfying the following boundary conditions:
\be\label{bc}
x(0)=\pm\frac{\ell}{2}\,,\qquad v(0)=t\,.
\ee
The area of this surface is given by the following functional:
\be\label{larangian}
\mathrm{Area}(\Gamma_A)\equiv\mathcal{A}(t)=\int_{0}^{z_*}dz\,\mathcal{L}\,,\qquad\mathcal{L}\equiv\frac{A_{\Sigma}}{z^{d-1}}\sqrt{x'^2-f(v,z)v'^2-2v'}\,,
\ee
where $A_{\Sigma}=2\ell_\perp^{d-2}$ is area of two $(d-2)-$dimensional hyperplanes. The constant $z_*$ here is defined through $x(z_*)=0$.
\item A $(d-1)-$dimensional ball of radius $R$, specified by
\begin{equation}
r^2\equiv\sum_i x_i^2\leq R\,.
\end{equation}
In this case it is convenient to write the $d\vec{x}^2$ in (\ref{vaid1}) in spherical coordinates:
\be
d\vec{x}^2=dr^2+r^2d\Omega^2_{d-2}\,.
\ee
The corresponding extremal surface $\Gamma_A$ is invariant under rotations. Therefore, without loss of generality,
we can parameterize it with two functions, $r(z)$ and $v(z)$, satisfying the following boundary conditions:
\be\label{bc2}
r(0)=R\,,\qquad v(0)=t\,.
\ee
The area of this surface is given by the following functional:
\be\label{larangian2}
\mathcal{A}(t)=\int_{0}^{z_*}dz\,\mathcal{L}\,,\qquad\mathcal{L}\equiv \frac{A_{\Sigma}r^{d-2}}{R^{d-2}z^{d-1}}\sqrt{r'^2-f(v,z)v'^2-2v'}\,,
\ee
where $A_{\Sigma}=2\pi^{\frac{d-1}{2}} R^{d-2}/\Gamma[\frac{d-1}{2}]$ is area of a $(d-2)$-dimensional spherical cap of radius $R$. The constant $z_*$ here is defined through $r(z_*)=0$.
\end{itemize}
We could go on and derive the equations of motion coming from (\ref{larangian}) and (\ref{larangian2}). However, these equations are generally highly non-linear so in practice one must proceed numerically. Our goal here will be to develop perturbative techniques in order to extract the explicit time dependence in various regimes of interest.

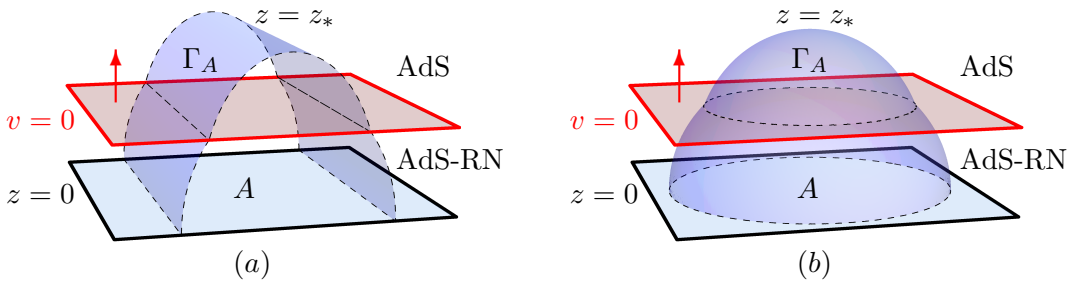
\begin{figure}[t!]
\begin{center}
\hspace{-0.65cm}\includegraphics[angle=0,width=0.55\textwidth]{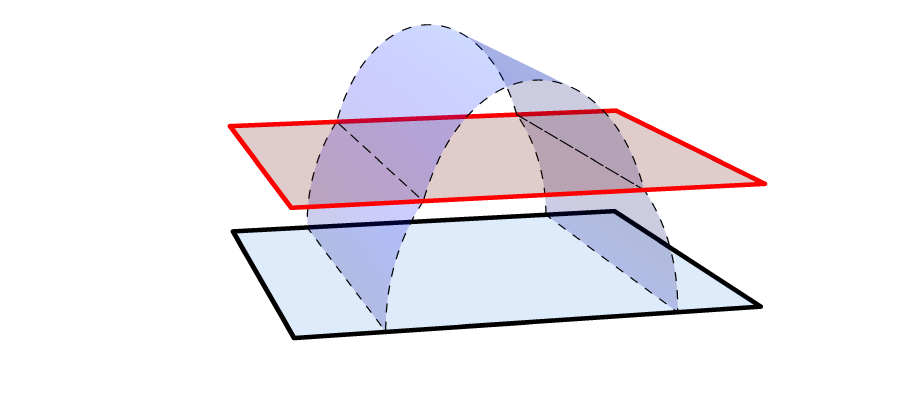}\hspace{-1.4cm} \includegraphics[angle=0,width=0.55\textwidth]{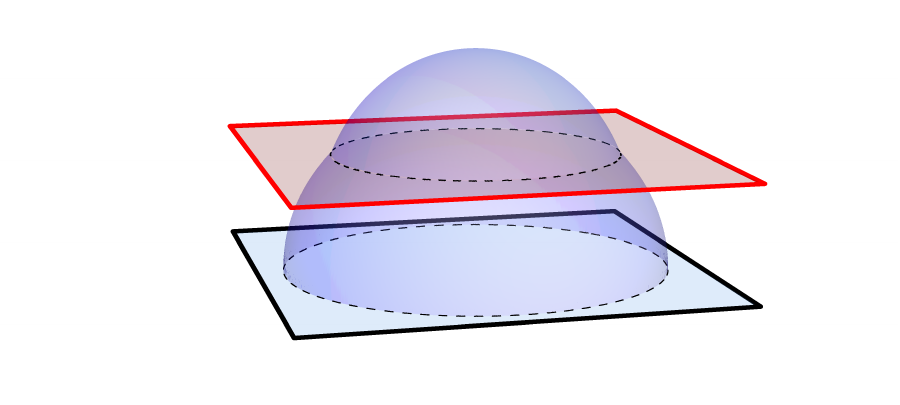}
\begin{picture}(0,0)
\put(-421,60){{\small {\color{red}$v=0$}}}
\put(-210,60){{\small {\color{red}$v=0$}}}
\put(-125,34){{\small $A$}}
\put(-336,34){{\small $A$}}
\put(-64,80){{\small AdS}}
\put(-275,80){{\small AdS}}
\put(-64,45){{\small AdS-RN}}
\put(-275,45){{\small AdS-RN}}
\put(-210,32){{\small $z=0$}}
\put(-421,32){{\small $z=0$}}
\put(-133,100){{\small $z=z_*$}}
\put(-328,100){{\small $z=z_*$}}
\put(-127,83){{\small $\Gamma_A$}}
\put(-355,83){{\small $\Gamma_A$}}
\put(-125,6){{\small $(b)$}}
\put(-336,6){{\small $(a)$}}
\thicklines
\put(-380,70){{\color{red}\vector(0,20){20}}}
\put(-169,70){{\color{red}\vector(0,20){20}}}
\end{picture}
\end{center}
\vspace{-0.7cm}
\caption{\small Extremal area surfaces in a thin shell Vaidya geometry for two different geometries: $(a)$ the strip and $(b)$ the ball. The shell (depicted in red) moves at the speed of light and eventually collapses into a black hole. The entanglement entropy of region $A$ grows as time evolves until the corresponding extremal surface $\Gamma_A$ grazes the shell at $v=0$. From this point on the whole surface lies entirely in the AdS-RN portion of the geometry so the entanglement entropy saturates to its final value.}
\label{figthinshell}
\end{figure}

Before doing so, let us discuss the thin shell regime, where $f(v,z)$ is given in terms of a step function as in (\ref{fstep}). The shell itself is located at $v=0$ and is moving towards the interior of the bulk. The regions $v<0$ and $v>0$ correspond to a pure AdS geometry and an AdS-RN black hole, respectively. A pictorial representation of the situation is given in Figure \ref{figthinshell}. One way to proceed is to
consider the regions $v<0$ and $v>0$ independently and then match the solutions across the shell, see e.g. \cite{Liu:2013iza,Liu:2013qca}. However, the analytical solution for $v>0$ is not known exactly so in practice one ends up expanding the solutions and picking up the relevant leading contributions. In particular, the work of \cite{Liu:2013iza,Liu:2013qca}
focused on the limit of large subsystems, where the main contribution comes from the near horizon portion of the geometry.
Here, we will consider a different approximation technique that is valid in the opposite regime, namely, for small subsystems.

\subsection{Perturbative expansion for small subsystems}

Besides the theoretical motivation presented in Section \ref{2dsec}, understanding the different analytical corners of the thermalization process is also interesting from a phenomenological point of view. One practical motivation is to shed light on the fast equilibration of the Quark Gluon Plasma (QGP), produced at ultra-relativistic heavy-ion collision experiments such as RHIC and LHC. In \cite{Caceres:2012em,Caceres:2014pda} it was noticed that in the limit of small subsystems, for near-thermal quenches ($T\gg\mu$) the saturation time \emph{decreases} with increasing chemical potential and thus the systems thermalizes faster. On the other hand, as we increase the size of the entangling region (in comparison to $1/T$) this behavior becomes less pronounced and eventually the saturation time starts increasing with the increase of chemical potential indicating that different physics take place at the two regimes of thermalization. Of course, these conclusions were based entirely on numerical calculations. We would like to understand this behavior better, using an appropriate approximation scheme.

In order to compute the leading behavior of the entanglement entropy we proceed in the following way. Consider the
functional $\mathcal{L}[\phi(z);\lambda]$ for the extremal surfaces, where $\phi(z)$ denote collectively the set of embedding functions, $\{x(z),v(z)\}$ for the strip or $\{r(z),v(z)\}$ for the ball, and $\lambda$ is a dimensionless parameter in which the perturbation will be carried out, i.e. $\lambda\ll1$. We can expand both $\mathcal{L}$ and $\phi(z)$ as follows:
\be
\begin{split}
\mathcal{L}[\phi(z);\lambda]&=\mathcal{L}^{(0)}[\phi(z)]+\lambda\mathcal{L}^{(1)}[\phi(z)]+\mathcal{O}(\lambda^2)\,,\\
\phi(z)&=\phi^{(0)}(z)+\lambda \phi^{(1)}(z)+\mathcal{O}(\lambda^2)\,.
\end{split}
\ee
In principle, the functions $\phi^{(n)}(z)$ could be obtained by solving the equations of motion order by order in $\lambda$. However, these equations are in general highly non-linear so in practice it is very difficult (and in most cases impossible) to obtain analytic results.
The key observation is that at first order in $\lambda$,\footnote{To our knowledge, this observation was first made in \cite{Hung:2011ta}.}
\be\label{arealambdaonshell}
\begin{split}
&\mathcal{A}_{\text{on-shell}}[\phi(z)]=\int dz\,\mathcal{L}^{(0)}[\phi^{(0)}(z)]+\lambda\int dz\,\mathcal{L}^{(1)}[\phi^{(0)}(z)]\\
&\qquad\qquad\qquad\,\,\,+\lambda \int dz\, \phi^{(1)}_i(z)\left[\cancel{\frac{d}{dz}\frac{\partial\mathcal{L}^{(0)}}{\partial \phi_i'(z)}-\frac{\partial\mathcal{L}^{(0)}}{\partial \phi_i(z)}}\right]_{\phi^{(0)}}\!\!\!+\cdots
\end{split}
\ee
Therefore, we only need $\phi^{(0)}(z)$ to obtain the first correction to the area. In our particular case, the expansion parameter is taken to be $\lambda\sim(\T \ell)^n$ (for some $n>1$), where $\ell$ is the characteristic length of the entangling region. Now, according to the UV/IR connection \cite{uvir1,uvir2,uvir3}, the bulk coordinate $z$ maps into a length scale in the boundary theory. In particular, since the extremal surface reach a maximum depth of $z_*$, then its natural to assume that $\ell\sim z_*$. On the other hand, the effective temperature is related to the inverse of the apparent horizon $\T\sim 1/z_H$ so, from the bulk perspective, having $\T \ell\ll1$ is equivalent to $z_*/ z_H\ll1$. Fortunately, in order to study this limit we just need the near boundary region, which is nothing but AdS plus small corrections. In the exact limit $\lambda\to0$ we expect to recover the embedding in pure AdS, which is known analytically.

\subsection{Explicit computation at leading order\label{pertcomp}}

\subsubsection{The strip}

Let us now make the above derivation more explicit. Since $z_*$ is actually the upper limit of integration in (\ref{arealambdaonshell}), we can first change to a new radial coordinate $y=z/z_*\in[0,1]$. The combination $z_*/z_H$ appears only in $f(v,z)$, which can now be expanded as
\be\label{metricexpansion}
f(v,y)=1-\theta(v)\,\varepsilon\, y^d\left(\frac{z_*}{z_H}\right)^d+\mathcal{O}\left(\frac{z_*}{z_H}\right)^{2(d-1)}\,.
\ee
At zeroth order in $z_*/z_H$ we get $f(v,z)=1$ and the spacetime is pure AdS, as expected. The leading correction is of order $(z_*/z_H)^d$ so in the field theory we expect corrections in $\lambda\sim(\T \ell)^d$. Expanding the area functional for the strip (\ref{larangian}), and going back to the original $z$ variable, it follows that
\be
\mathcal{L}^{(0)}=\frac{A_\Sigma}{z^{d-1}}\sqrt{x'^2-v'^2-2v'}\,,\qquad\;\;\;\mathcal{L}^{(1)}=\frac{\varepsilon A_\Sigma}{2z_H^d}\frac{z v'^2 \theta(v)}{\sqrt{x'^2-v'^2-2v'}}\,.
\ee
We also need the embedding functions at zeroth order $\{x(z),v(z)\}$. For $f(v,z)=1$ the spacetime is static so all extremal surfaces lie on a constant-$t$ slice, $t(z)=t$. Equation (\ref{efcoords}) then yields
\be\label{vofzeq}
v(z)=t-z\,.
\ee
Plugging (\ref{vofzeq}) back into $\mathcal{L}^{(0)}$ we obtain the standard area functional in empty AdS, which has the known solution \cite{Hubeny:2012ry}
\be\label{embAdS}
x(z)=\frac{\ell}{2}-\frac{z_*}{d}\left(\frac{z}{z_*}\right)^d\,\!_2F_1\left[\frac{1}{2},\frac{d}{2(d-1)},\frac{3d-2}{2(d-1)},\left(\frac{z}{z_*}\right)^{2(d-1)}\right]\,,
\ee
with
\be\label{ellzst}
\ell=\frac{2 \sqrt{\pi }\Gamma[\frac{d}{2(d-1)}]z_*}{\Gamma[\frac{1}{2(d-1)}]}\,.
\ee
The zeroth order contribution to the area is time-independent and includes all UV divergences. Here we are interested in the time-dependent part only, so we will focus on the quantity
\be\label{deltaSeq}
\Delta S_A(t) = \frac{\Delta \mathcal{A}(t)}{4G_N^{(d+1)}}=\frac{1}{4G_N^{(d+1)}}\int dz\,\mathcal{L}^{(1)}[\phi^{(0)}(z)]+\cdots\,,
\ee
where $\Delta\mathcal{A}(t)\equiv\mathcal{A}(t)-\mathcal{A}_{\text{AdS}}$ and the dots denote higher order terms in $\lambda$.\footnote{In Appendix \ref{nexttoleading} we compute the first sub-leading term in this expansion.} Note that with this subtraction $\Delta S_A(t)$ naturally starts from zero in the infinite past. Evaluating the leading order term of (\ref{deltaSeq}) on shell leads to
\be\label{deltaSon}
\Delta S_A(t) =\frac{\varepsilon A_\Sigma}{8G_N^{(d+1)}z_H^d}\int_0^{z_*}dz\, \theta (t-z) z \sqrt{1-(z/z_*)^{2(d-1)}}\,.
\ee
In order to evaluate this integral it is convenient to define a new variable $\xi=t-z$. With this substitution, the integral in (\ref{deltaSon}) becomes
\be\label{intdef}
I=\int_{t-z_*}^{t} d\xi\, \theta (\xi) (t-\xi) \sqrt{1-[(t-\xi)/z_*]^{2(d-1)}}\,.
\ee
Let us consider the following three cases, $(i)$ $t<0$, $(ii)$ $0<t<z_*$ and $(iii)$ $z_*<t$:\\
$(i)$ Since both limits are negative and $\theta(\xi<0)=0$, then
\be
I=0\,.
\ee
$(ii)$ The lower limit is negative so we can replace it by zero:
  \bea
  I&=&\int_0^{t} d\xi\,(t-\zeta ) \sqrt{1-[(t-\xi)/z_*]^{2(d-1)}}=\int_0^{t}dz\,z \sqrt{1-(z/z_*)^{2(d-1)}}\,,\\
  &=&\frac{t^2}{d+1} \left\{\sqrt{1-\left(\frac{t}{z_*}\right)^{2(d-1)}}+\frac{d-1}{2}\!\,_2F_1\left[\frac{1}{2},\frac{1}{d-1},\frac{d}{d-1},\left(\frac{t}{z_*}\right)^{2(d-1)}\right]\right\}.
  \eea
$(iii)$ Since both limits are positive and $\theta(\xi>0)=1$, we get:
\bea
I=\int_0^{z_*}dz\,z \sqrt{1-(z/z_*)^{2(d-1)}}=\frac{\sqrt{\pi}\Gamma[\frac{1}{d-1}]z_*^2}{2(d+1)\Gamma [\frac{d+1}{2(d-1)}]}\,.
\eea
Notice that this last expression is independent of time, so in this approximation the saturation time is given by
\be\label{tsatdef}
t_{\text{sat}}=z_*=\frac{\Gamma[\frac{1}{2(d-1)}]\ell}{2\sqrt{\pi}\Gamma[\frac{d}{2(d-1)}]}\,.
\ee
Altogether, the leading correction to the entanglement entropy can be expressed as
\be\label{deltaSfinal}
\Delta S_A(t)=\Delta S_{\text{eq}}\big\{[\theta(t)-\theta(t-t_{\text{sat}})]\mathcal{F}(t/t_{\text{sat}})+ \theta(t-t_{\text{sat}})\big\}\,,
\ee
where $\Delta S_{\text{eq}}$ is the final value of the entropy,
\be
\Delta S_{\text{eq}}=\frac{\sqrt{\pi}\Gamma[\frac{1}{d-1}]z_*^2A_\Sigma\varepsilon }{16(d+1)\Gamma[\frac{d+1}{2(d-1)}]z_H^dG_N^{(d+1)}}\,,
\ee
and $\mathcal{F}$ is given by:
\be\label{defFfun}
\mathcal{F}(x)=\frac{2\Gamma[\frac{d+1}{2(d-1)}]x^2}{\sqrt{\pi}\Gamma[\frac{1}{d-1}]} \left[\sqrt{1-x^{2(d-1)}}+\tfrac{d-1}{2}\!\,_2F_1\left(\tfrac{1}{2},\tfrac{1}{d-1},\tfrac{d}{d-1},x^{2(d-1)}\right)\right].
\ee
By definition the function $\mathcal{F}$ satisfies that $\mathcal{F}(0)=0$ and $\mathcal{F}(1)=1$, so in this range its average rate of change is $\langle d\mathcal{F}(x)/dx\rangle=1$. With this result, we can now compute the instantaneous rate of entanglement growth,\footnote{
A comment on the normalization of (\ref{insrate}) is in order: similar to (\ref{defseq}), here $s_{\text{eq}}=\Delta S_{\text{eq}}/V_A$ refers to the equilibrium \emph{entanglement entropy} (rather than thermal entropy) after the quench in a volume $V_A$. For small subsystems, the entanglement entropy of excited states obeys a first-law like relation $\Delta E_A= T_{\text{ent}}\Delta S_A$, where $T_{\text{ent}}$ is the so-called entanglement temperature \cite{Bhattacharya:2012mi,Allahbakhshi:2013rda}. Therefore, in this limit $s_{\text{eq}}=S_{\text{eq}}/V_A=T_{\text{ent}}^{-1}\mathcal{E}$, where $\mathcal{E}$ is the energy density of the final state.}
\be\label{insrate}
\mathfrak{R}(t)=\frac{1}{s_{\text{eq}}A_\Sigma}\frac{d(\Delta S_A)}{dt}=\frac{V_A}{A_\Sigma t_{\text{sat}}}\frac{d\mathcal{F}}{dx}=\frac{2 (d+1)\Gamma[\frac{d}{2(d-1)}] \Gamma[\frac{d+1}{2(d-1)}]}{\Gamma [\frac{1}{d-1}]\Gamma[\frac{1}{2(d-1)}]}x\sqrt{1-x^{2 (d-1)}}\,.
\ee
where $x=t/t_{\text{sat}}$ and the time-averaged entanglement velocity:
\be\label{vavgs}
v_E^{\text{avg}}=\langle\mathfrak{R}(t)\rangle=\frac{V_A}{A_\Sigma t_{\text{sat}}}=
\frac{\sqrt{\pi}\Gamma[\frac{d}{2(d-1)}]}{\Gamma[\frac{1}{2(d-1)}]}=\begin{cases}
\displaystyle 1\, , & \displaystyle  \quad d=2 \,,\\[1ex]
 \displaystyle 0.5991\, , & \displaystyle  \quad d=3 \,,\\[1ex]
  \displaystyle 0.4312\, ,
  &\displaystyle  \quad d=4\,,\\[1ex]
  \displaystyle 0\, ,
  &\displaystyle  \quad d\to\infty\,.
\end{cases}
\ee
In the above, we have used the expressions for the strip, $V_A=\ell_\perp^{d-2}\ell$ and $A_\Sigma=2\ell_\perp^{d-2}$.

\begin{figure}[t!]
$$
\begin{array}{cc}
  \includegraphics[angle=0,width=0.43\textwidth]{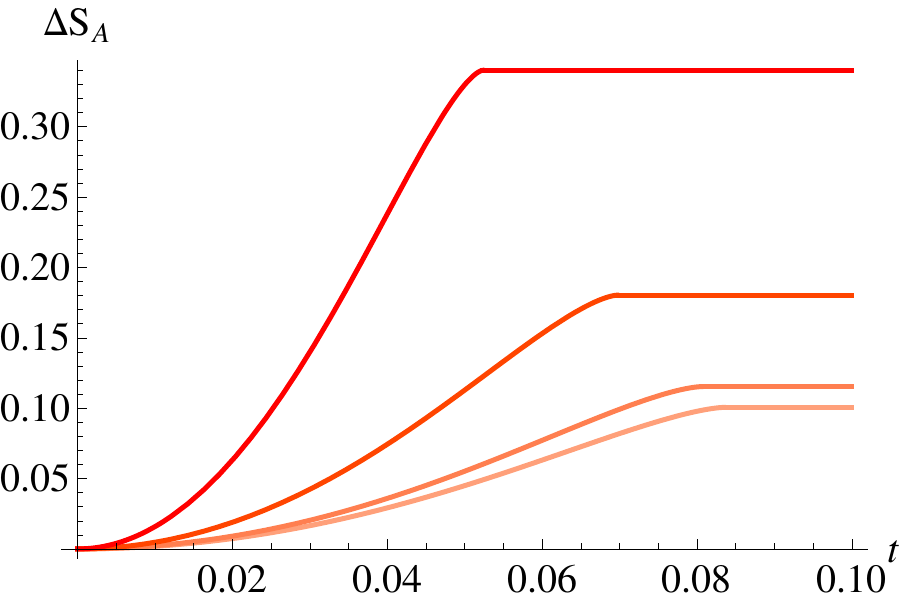} \quad &\quad \includegraphics[angle=0,width=0.43\textwidth]{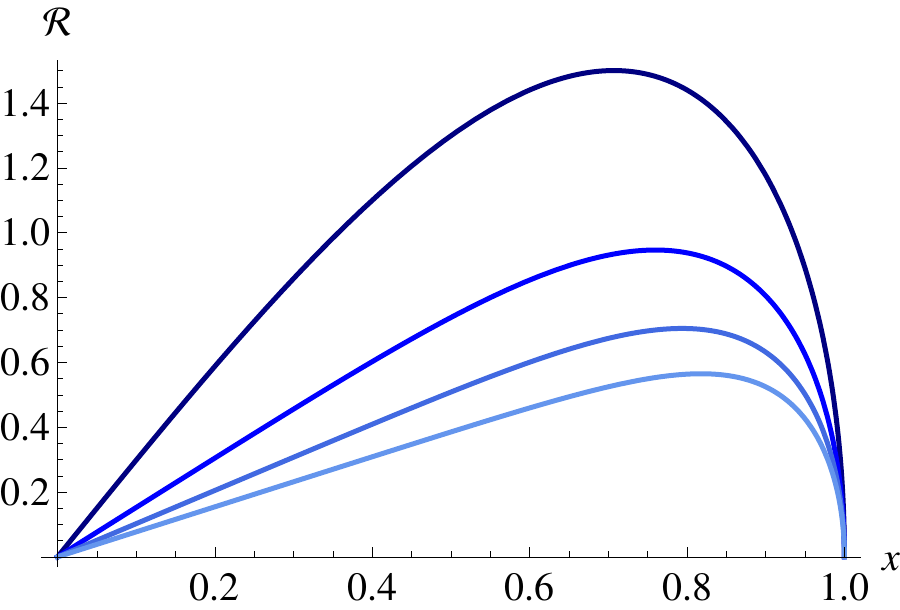}\\
  (a) \quad  & \quad (b)
\end{array}
$$
\vspace{-0.7cm}
\caption{\small $(a)$ Evolution of entanglement entropy for a strip in $d=3$ and $\mu/T=\{0,2,5,10\}$ from bottom to top, respectively. For the plots we have fixed $\ell\T=10^{-1}$ so that the approximation is valid and we have set the overall factor $A_\Sigma/4G_N^{(d+1)}=1$. According to (\ref{tsatdef}), the saturation time scales as $t_{\text{sat}}\sim\ell$ which, for our particular choice of parameters, translates into $t_{\text{sat}}\sim1/\T$. Both, the differences in final entropies and saturation times become more pronounced as we increase the number of dimensions, but the behavior is qualitatively similar. In $(b)$ we plot the instantaneous rate of growth for $\mathfrak{R}(x)$ for $d=\{2,3,4,5\}$ from top to bottom, respectively. We observe that the maximum rate growth only exceed the speed of light for $d=2$, and decreases as we increase the number of dimensions.}
\label{EEVaidya}
\end{figure}

Before analyzing in detail the different regimes of (\ref{deltaSfinal}), let us first briefly comment on some generalities. In Figure \ref{EEVaidya} $(a)$ we plot the evolution of entanglement entropy for some sample parameters. In general, we observe a qualitatively similar behavior for the entanglement entropy as the numerical results of \cite{Caceres:2012em,Caceres:2014pda}. However, at this level of approximation $t_{\text{sat}}\sim\ell$ so it is clear that our result does not capture the non-monotonic behavior with respect to $\mu/T$ obtained in these references. In particular, for our plots we have chosen to keep $\ell\T=$ fixed so it is clear that the change in saturation time is entirely due to the variation of $\T$ as we increase $\mu/T$, which is always monotonic. We will come back to this point in Section \ref{correction}, where we explicitly compute the leading corrections to the $t_{\text{sat}}$. In particular, we will show that the first correction is enough to observe the expected behavior reported in \cite{Caceres:2012em,Caceres:2014pda}.
In $(b)$ we plot the instantaneous rate of growth (\ref{insrate}) as a function of $x=t/t_{\text{sat}}$. For $d=2$ we get exactly the same curve as in Figure \ref{EE2db}, with a maximum of $\text{max}[\mathfrak{R}(t)]=3/2$, so we can view it as a consistency check of our perturbative method. For $d\geq3$ the maximum rate is always below the speed of light and decreases monotonically as we increase the number of dimensions. We will discuss this point in more detail in Section \ref{quasilinear}.

\subsubsection{The ball}

The computation for the ball is very similar to the case of the strip, so we will only sketch the main few steps. Expanding the area functional (\ref{larangian2}) it follows that
\be
\mathcal{L}^{(0)}=\frac{A_\Sigma r^{d-2}}{R^{d-2}z^{d-1}}\sqrt{r'^2-v'^2-2v'}\,,\qquad\;\;\;\mathcal{L}^{(1)}=\frac{\varepsilon A_\Sigma}{2R^{d-2}z_H^d}\frac{zr^{d-2} v'^2 \theta(v)}{\sqrt{r'^2-v'^2-2v'}}\,.
\ee
We also need the embedding functions $\{r(z),v(z)\}$ in pure AdS. For the case of the ball $v(z)$ is still given by (\ref{vofzeq}) but $r(z)$ now takes the form of a spherical cap \cite{Hubeny:2012ry}
\be\label{sphericalsol}
r(z)=\sqrt{z_*^2-z^2}\,,\qquad\qquad\qquad R=z_*\,.
\ee
Again, we are interested in the difference of entanglement with respect to pure AdS, so we focus on the $\mathcal{L}^{(1)}$ piece only. Evaluating this term on shell leads to:
\be\label{deltaSon2}
\Delta S_A(t)=\frac{\varepsilon A_\Sigma z_*^{d-2}}{8G_N^{(d+1)}R^{d-2}z_H^d}\int_0^{z_*}dz\, \theta (t-z) z \left[1-\left(z/z_*\right)^2\right]^{\frac{d-1}{2}}\,,
\ee
which resembles (\ref{deltaSon}) and can be evaluated in a similar way. The upshot of the calculation is
\be\label{deltaSfinal2}
\Delta S_A(t)=\Delta S_{\text{eq}}\big\{[\theta(t)-\theta(t-t_{\text{sat}})]\mathcal{G}(t/t_{\text{sat}})+ \theta(t-t_{\text{sat}})\big\}\,,
\ee
where
\be
t_{\text{sat}}=z_*=R\,,
\ee
\be
\Delta S_{\text{eq}}=\frac{R^2A_\Sigma\varepsilon }{8(d+1)z_H^dG_N^{(d+1)}}\,,
\ee
and $\mathcal{G}$ is given by:
\be\label{defGfun}
\mathcal{G}(x)=1-\left(1-x^2\right)^{\frac{d+1}{2}}\,.
\ee
We can also compute the instantaneous rate of change of the entanglement growth,
\be\label{insrate2}
\mathfrak{R}(t)=\frac{V_A}{A_\Sigma t_{\text{sat}}}\frac{d\mathcal{F}}{dx}=\frac{(d+1)}{(d-1)}x \left(1-x^2\right)^{\frac{d-1}{2}}\,.
\ee
where $x=t/t_{\text{sat}}$ and the time-averaged entanglement velocity:
\be\label{vavgb}
v_E^{\text{avg}}=\frac{V_A}{A_\Sigma t_{\text{sat}}}=
\frac{1}{d-1}=\begin{cases}
\displaystyle 1\, , & \displaystyle  \quad d=2 \,,\\[1ex]
 \displaystyle \frac{1}{2}\, , & \displaystyle  \quad d=3 \,,\\[1ex]
  \displaystyle \frac{1}{3}\, ,
  &\displaystyle  \quad d=4\,,\\[1ex]
  \displaystyle 0\, ,
  &\displaystyle  \quad d\to\infty\,.
\end{cases}
\ee
In the above, we have used the expressions for the ball, $V_A=2\pi^{\frac{d-1}{2}} R^{d-1}/\Gamma[\frac{d-1}{2}](d-1)$ and $A_{\Sigma}=2\pi^{\frac{d-1}{2}} R^{d-2}/\Gamma[\frac{d-1}{2}]$.

\begin{figure}[t!]
$$
\begin{array}{cc}
  \includegraphics[angle=0,width=0.43\textwidth]{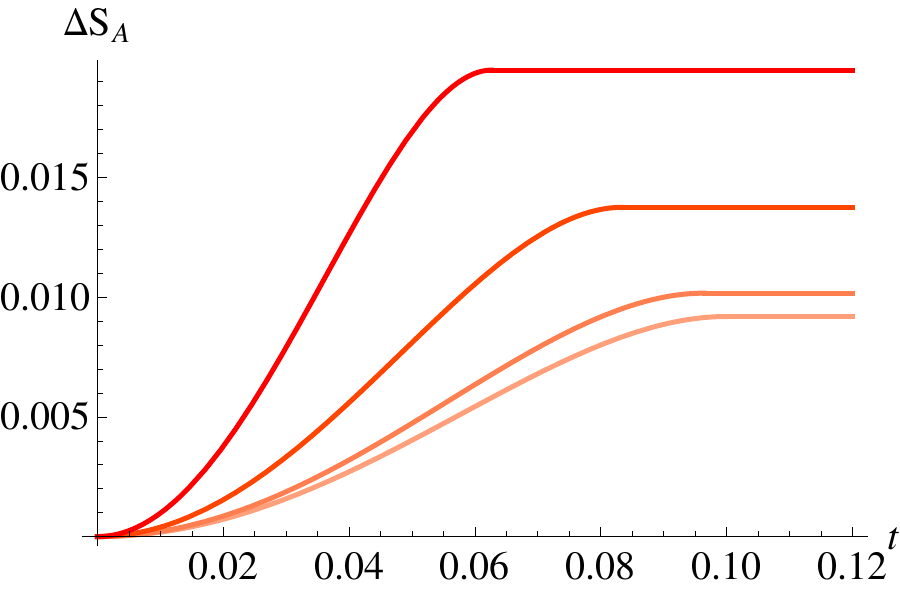} \quad &\quad \includegraphics[angle=0,width=0.43\textwidth]{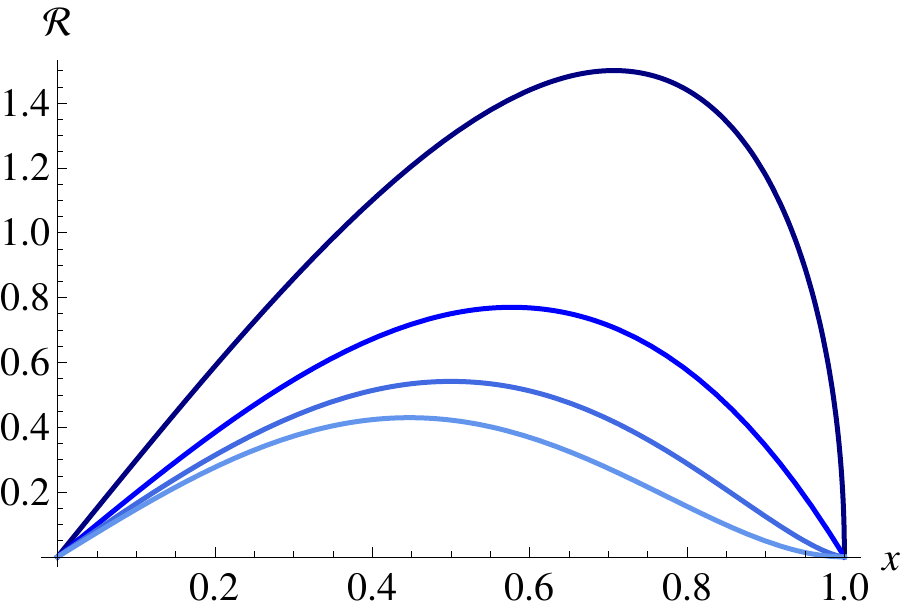}\\
  (a) \quad  & \quad (b)
\end{array}
$$
\vspace{-0.7cm}
\caption{\small $(a)$ Evolution of entanglement entropy for a ball in $d=3$ and $\mu/T=\{0,2,5,10\}$ from bottom to top, respectively. For the plots we have set $R\T=10^{-1}$ and $A_\Sigma/4G_N^{(d+1)}=1$. In $(b)$ we plot the instantaneous rate of growth for $\mathfrak{R}(x)$ for $d=\{2,3,4,5\}$ from top to bottom, respectively. Again, the maximum rate growth only exceed the speed of light for $d=2$.}
\label{EEVaidya2}
\end{figure}

The behavior of these observables is qualitatively similar to the case of the strip. In Figure \ref{EEVaidya2} we plot the entanglement growth and the instantaneous rate of change for some sample parameters. For the entanglement growth curves in $(a)$ we have keep $R\T$ fixed so the saturation time is monotonic in $\mu/T$. We will compute the first correction to $t_{\text{sat}}$ in Section \ref{correction}. From the curves in $(b)$ we observe that: $i)$ the instantaneous rate of growth does not exceed the speed of light for $d\geq3$ and $ii)$ $\mathfrak{R}\to0$ as $x\to1$ so the approach to saturation is continuous. All these behaviors are likely to hold for more general entangling surfaces.

\section{Regimes of thermalization\label{secregimes}}

Let us now analyze in more detail our results for the strip (\ref{deltaSfinal}) and the ball (\ref{deltaSfinal2}) specializing to the different regimes of thermalization. Specifically, we will focus on three distinct regimes: the initial quadratic growth, an intermediate quasi-linear growth and the saturation.

\subsection{Initial quadratic growth\label{initialg}}

The initial growth regime is dominated by the behavior of $\mathcal{F}(x)$ or $\mathcal{G}(x)$ for $x\ll1$. Expanding these functions we get
\be
\mathcal{F}(x)=\frac{(d+1) \Gamma[\frac{d+1}{2(d-1)}]}{\sqrt{\pi}\Gamma[\frac{1}{d-1}]}x^2+\mathcal{O}(x^{2d})\,,
\ee
and
\be
\mathcal{G}(x)=\frac{1}{2} (d+1) x^2+\mathcal{O}(x^{4})\,,
\ee
respectively. In both cases, the early time growth of the entanglement is given by
\be \label{quda}
\Delta S_A (t) = {A_\Sigma\varepsilon \over 16 z_H^d G_N^{(d+1)}}  t^2 + \cdots \ .
\ee
The fact that equation (\ref{quda}) applies for both, the strip and the ball, suggests a universal behavior at early times; we will comment more on this below. We can also express this result in terms of the physical data $T$ and $\mu$. The general expression is a little cumbersome so, for the sake of simplicity, we will only consider the following two limits:
\begin{enumerate}
\item Near-thermal quenches ($T\gg\mu$):
\be\label{igthermal}
\Delta S_A (t) = \frac{A_\Sigma}{16G_N^{(d+1)}}\left(\frac{4 \pi  T}{d}\right)^d\left(1+\frac{d^2(d-2)}{16 \pi ^2}\left(\frac{\mu }{T}\right)^2+\cdots\right)t^2+\cdots\,.
\ee
\item Near-extremal quenches ($T\ll \mu$):
\be\label{igextremal}
\Delta S_A (t) =\frac{A_\Sigma}{16 G_N^{(d+1)}}\frac{2 (d-2)^{d-1} \mu^d}{d^{d/2}(d-1)^{d/2-1}}\left(1+\frac{2\pi d^{1/2}}{(d-1)^{1/2}} \frac{T}{\mu }+\cdots\right)t^2 + \cdots\,.
\ee
This last result includes the extremal case, for which $T=0$.
\end{enumerate}
We can also verify that our results agree with the ones presented in \cite{Liu:2013iza,Liu:2013qca} for large subsystems. This is another clear indication that in the early growth regime the evolution of entanglement is independent of the size and shape of the entangling region as long as $t\ll t_{\text{sat}}$. Furthermore, the absence of additional geometric quantities such as $\ell$ or $R$ in the expression (\ref{quda}) (besides $A_\Sigma$ itself) implies that the quadratic growth behavior $\Delta S_A (t)\sim t^2$ may be entirely fixed by symmetries (more specifically, conformal symmetry). We will confirm these claims explicitly in Section \ref{earlyuniversal}.

\subsection{Quasi-linear growth\label{quasilinear}}

For large regions, entanglement entropy exhibits a universal intermediate regime \cite{Liu:2013iza,Liu:2013qca}
\be\label{lgrowthb}
\Delta S_A (t)=v_E s_{\text{eq}} A_\Sigma t\,,\qquad t_{\text{sat}}\gg t\gg t_{\text{loc}}\,,
\ee
where $s_{\text{eq}}$ is the entropy density of the final state, $s_{\text{eq}}=\Delta S_{\text{eq}}/V_A$, and $v_E$ is the so called ``tsunami velocity''. The local equilibrium scale $t_{\text{loc}}$ is given by the position of the horizon $t_{\text{loc}}\sim z_H$, which can be rewritten as $t_{\text{loc}}\sim1/\T$. Of course, in this limit the physics differs drastically from the regime we are focusing on: entanglement entropy approaches the thermodynamic entropy and the main contribution to the extremal surfaces comes from the interior of the bulk geometry. Another crucial difference is that for small subsystems we cannot really talk about ``local equilibrium'' before the entanglement entropy reaches saturation. We will, nevertheless, attempt to make a comparison between the two regimes and point out the main similarities and differences.

Let us begin by reviewing more explicitly the results of \cite{Liu:2013iza,Liu:2013qca} for the charged case. In these papers the authors found that for large subsystems
\be\label{veofu}
v_E=\sqrt{\frac{d}{d-2}}\left(\left(1-\frac{d\,u}{2(d-1)}\right)^{\frac{2(d-1)}{d}}-\left(1-u\right)\right)^{\frac{1}{2}}\,,\qquad u\equiv\frac{4\pi z_HT}{d}=\frac{T}{\T}\,.
\ee
The parameter $u$ lies in the range $0\leq u\leq1$ and decreases monotonically from its Schwarzschild
value $u=1$ to $u=0$, as the $\mu/T$ is increased from zero to infinity. Given the dependence of (\ref{veofu}) on $u$
this implies that turning on a nonzero chemical potential always slows down the
evolution. Let us study more closely the small-$\mu/T$ and large-$\mu/T$ limits of (\ref{veofu}). For small $\mu/T$ we get that, at leading order
\be\label{veholo}
v_E=\sqrt{\frac{d}{d-2}} \left(\frac{d-2}{2(d-1)}\right)^{\frac{d-1}{d}}=\begin{cases}
\displaystyle 1\, , & \displaystyle  \quad d=2 \,,\\[1ex]
 \displaystyle 0.6874\, , & \displaystyle  \quad d=3 \,,\\[1ex]
  \displaystyle 0.6204\, ,
  &\displaystyle  \quad d=4\,,\\[1ex]
  \displaystyle 1/2\, ,
  &\displaystyle  \quad d\to\infty\,,
\end{cases}
\ee
while for large $\mu/T$ (and $d\geq3$) we get
\be\label{veholomu}
v_E=\frac{2\pi}{d-2}\left(\frac{T}{\mu}\right)\to0\,.
\ee
The fact that $v_E\to0$ when the quench approaches extremality implies that the linear growth regime no longer exists.
This was indeed observed numerically in \cite{Albash:2010mv}. In this case, the linear growth regime is replaced by a logarithmic growth regime.

Let us now go back to the case of small subsystems. Our results for the strip (\ref{deltaSfinal}) and the ball (\ref{deltaSfinal2})
indicate that in this case the evolution is non-universal. More precisely, since the normalized rate of change $\mathfrak{R}(t)$ is different in these two cases, we can conclude that the equilibration process for small subsystems strongly depends on the shape of the entangling region. Moreover, since
the growth of entanglement is not strictly linear in either case so we cannot define a velocity in the sense of (\ref{lgrowthb}). Instead, we will define a quasi-linear regime based on the maximum rate of growth of the entanglement entropy:
\be\label{vedefinition}
v_E^{\text{max}}\equiv\text{max}[\mathfrak{R}(t)]=\frac{1}{s_{\text{eq}} A_\Sigma}\frac{d(\Delta S_A)}{d t}\bigg|_{t=t_\text{max}}\,.
\ee
A few comments are in order. First note that this would be natural way to define an analogue of the tsunami velocity $v_E$ since at $t=t_\text{max}$
\be\label{quasiregime}
\Delta S_A(t)-\Delta S_A(t_{\text{max}})=v_E^{\text{max}}s_{\text{eq}}A_\Sigma(t-t_{\max})+\mathcal{O}(t-t_{\max})^3\,,
\ee
so the quadratic corrections to the rate of change of the entanglement entropy vanish. However, since this linear behavior is \emph{instantaneous}
we argue that the heuristic  picture  for  the  entanglement growth  in terms of a wave propagating inwards from the boundary $\Sigma$ does not
hold in this regime. This is indeed expected, since for small subsystems the spread of entanglement takes place at timescales that are shorter in
comparison to the local equilibration scale $t_{\text{loc}}$. Second, the value of $v_E^{\text{max}}$ generally depends on the shape of the entangling region, so the equation (\ref{quasiregime}) is non-universal. For the strip, and at leading order in $\ell\T$ we find that
\be
v_E^{\text{max}}=\frac{\ell}{2t_{\text{sat}}}\frac{d\mathcal{F}}{d x}\bigg|_{x=x_\text{max}}\,,
\ee
where $\mathcal{F}(x)$ is given in (\ref{defFfun}). The first derivative of $\mathcal{F}$ is given by
\be
\frac{d\mathcal{F}}{d x}=\frac{2 (d+1) \Gamma[\frac{d+1}{2(d-1)}]}{\sqrt{\pi} \Gamma [\frac{1}{d-1}]}x\sqrt{1-x^{2 (d-1)}}\,.
\ee
It first increases linearly, reach a maximum at some $x_{\text{max}}$ and then decreases all the way to zero, at $x=1$. The maximum is attained at:
\be
\frac{d^2\mathcal{F}}{d x^2}=0\qquad\longrightarrow\qquad x_{\text{max}}=\frac{1}{d^{\frac{1}{2(d-1)}}}\,,
\ee
and is given by
\be
\frac{d\mathcal{F}}{d x}\bigg|_{x=x_\text{max}}=\frac{4(d-1)^{3/2} \Gamma[\frac{3d-1}{2(d-1)}]}{\sqrt{\pi } d^{\frac{d}{2(d-1)}} \Gamma [\frac{1}{d-1}]}\,.
\ee
The expression for $t_{\text{sat}}$ is given in (\ref{tsatdef}). Putting all together we find that for the strip
\be\label{vesamallL}
v^{\text{max}}_E=\frac{4(d-1)^{3/2} \Gamma [\frac{3 d-1}{2 (d-1)}] \Gamma [\frac{d}{2 (d-1)}]}{d^{\frac{d}{2 (d-1)}} \Gamma [\frac{1}{2 (d-1)}] \Gamma [\frac{1}{d-1}]}=\begin{cases}
 \displaystyle \frac{3}{2}\, , & \displaystyle  \quad d=2 \,,\\[1ex]
 \displaystyle 0.9464\, , & \displaystyle  \quad d=3 \,,\\[1ex]
  \displaystyle 0.7046\, ,
  &\displaystyle  \quad d=4\,,\\[1ex]
  \displaystyle \pi/d\to0
  \, ,
  &\displaystyle  \quad d\to\infty\,.
\end{cases}
\ee
We can follow similar same steps for the case of the ball. At the end of the computation, we find that in this case
\be\label{vesamallL}
v^{\text{max}}_E=\frac{(1+d)(d-1)^{\frac{d-3}{2}}}{d^{d/2}}=\begin{cases}
 \displaystyle \frac{3}{2}\, , & \displaystyle  \quad d=2 \,,\\[1ex]
 \displaystyle 0.7698\, , & \displaystyle  \quad d=3 \,,\\[1ex]
  \displaystyle 0.5413\, ,
  &\displaystyle  \quad d=4\,,\\[1ex]
  \displaystyle 1/\sqrt{ed}\to0\, ,
  &\displaystyle  \quad d\to\infty\,,
\end{cases}
\ee
giving a lower maximum rate in comparison to the strip. On the other hand, it is interesting that for small subsystems the maximum velocity $v_E^{\text{max}}$ (and more generally, the instantaneous rate $\mathfrak{R}(t)$) is \emph{independent} of $T$ and $\mu$, contrary to the large interval result (\ref{veofu}). Thus, we can say that $v_E^{\text{max}}$ is independent of the state, whereas $v_E$ is independent of the entangling region.
Comparing the two quantities, we can also observe that the maximum rate of change of entanglement entropy can be faster in the UV for $d\leq4$ ($d\leq3$ for the ball) as long as $\mu/T\ll1$, but it is generally slower in higher dimensions. For $\mu/T\gg1$ the maximum rate is always faster in the UV.

It is remarkable that $v^{\text{max}}_E$ can in some cases exceed the value of the tsunami velocity $v_E$, which had been previously proposed as an upper bound for the rate of change of the entanglement entropy \cite{Liu:2013iza,Liu:2013qca}. However, we should bear in mind that the physics in these two scenarios is completely different. Specifically, the bound proposed in \cite{Liu:2013iza,Liu:2013qca} seems to apply specifically to the growth of entanglement after local equilibration has been achieved, in the strict limit of large subsystems. More recently, the authors of \cite{Casini:2015zua,Hartman:2015apr} showed that $v_E$ is actually bounded by the speed of light, i.e. $v_E\leq1$, even though $v_E$ is not actually a physical velocity. Here, we argue that $v^{\text{max}}_E$ (and more generally $\mathfrak{R}(t)$) is not constrained by this bound, even though for holographic models the violation only appears for $(1+1)-$dimensional theories. On the other hand, it seems reasonable to assume that for general $d$, the \emph{total} equilibration time $t_{\text{sat}}$ must be at least the light-crossing time
of region $A$, so the average entanglement velocity $v^{\text{avg}}_E$ must be bounded by the speed of light, $v^{\text{avg}}_E\leq1$.
For the case of small subsystems this bound holds for both, the strip (\ref{vavgs}) and the ball (\ref{vavgb}). For large subsystems it is valid in general, given that in this limit $v_E\leq v^{\text{avg}}_E\leq1$ (e.g. for a strip $v_E=v^{\text{avg}}_E$ and the inequality is saturated, but for a ball $v_E< v^{\text{avg}}_E$). We believe that $v^{\text{avg}}_E$ represents a more honest comparison between entangling regions of different sizes. Indeed, if we compare the results of $v^{\text{avg}}_E$ for small subsystems (\ref{vavgs}), (\ref{vavgb}) with those for large subsystems (\ref{veholo}), (\ref{veholomu}) we can reach a more universal conclusion for the process of thermalization: in average the UV degrees of freedom equilibrate at a slower rate than the IR degrees of freedom when the evolution is governed by thermal fluctuations ($\mu/T\ll1$) but at a faster rate if the evolution is driven by quantum fluctuations ($\mu/T\gg1$). This conclusion is more robust than the one reached for $v^{\text{max}}_E$ because it is independent of the number of dimensions and the shape of the entangling region $A$.

\subsection{Approach to saturation\label{correction}}

For large subsystems, the authors of \cite{Liu:2013iza,Liu:2013qca} found that the equilibration of the entanglement entropy depends quite generally on
the shape of the entangling region, the spacetime dimension
$d$, and the final state. For the strip, in particular,
it was found that for general $d\geq3$ the transition is quite abrupt:
the first derivative of $\Delta S_A (t)$ is generally discontinuous at $t=t_{\text{sat}}$,
in analogy to a first-order phase transition. For small subregions,
this stage can be studied by expanding $\mathcal{F}(x)$ or $\mathcal{G}(x)$ around $x=1$.
For the strip we find that the saturation
resembles that of a continuous (second-order) phase transition with
\be
\Delta S_A (t)-S_{\text{eq}}\propto(t_{\text{sat}}-t)^{\gamma}\,,\qquad\qquad\gamma=\frac{3}{2}\,.
\ee
However, it differs from the mean-field behavior $\gamma=2$ of standard thermodynamic transitions. It is worth emphasizing that the phase transition observed for large subregions is due to an abrupt exchange of dominance of extremal surfaces at $t=t_{\text{sat}}$. The origin of this feature is well understood since the earlier numerical studies of \cite{Balasubramanian:2010ce,Balasubramanian:2011ur}: it is due to the multi-valuedness of $z_*(\ell)$ near the saturation time, which in turn leads to a swallow-tail behavior of the entanglement entropy. For small regions, however, the leading contributions come from the pure AdS embedding, which has a unique value of $z_*(\ell)$, regardless of the temporal evolution. We expect this multi-valuedness to appear at some point once we include higher order corrections in $\ell\T$.

The case of the ball is a little more subtle. In \cite{Liu:2013iza,Liu:2013qca} it was found that,
for $R\T\gg1$, the same discontinuous behavior also appears for $d=3$ as long as $\mu/T\gg1$.
On the other hand, for general $d\geq4$ the approach to saturation is continuous, and is
characterized by a nontrivial scaling exponent
\be\label{decay}
\Delta S_A (t)-S_{\text{eq}}\propto(t_{\text{sat}}-t)^{\gamma}\,,\qquad\qquad\gamma=\frac{d+1}{2}\,.
\ee
The same exponent applies for $d=2$, while for $d=3$ and $\mu/T\ll1$ it was found that
$\Delta S_A (t)-S_{\text{eq}}\propto(t_{\text{sat}}-t)^{2}\log(t_{\text{sat}}-t)$, marginally avoiding the mean-field exponent $\gamma=2$.\footnote{For the cases in which the saturation is
continuous, the authors of \cite{Liu:2013iza,Liu:2013qca} referred to the stage
prior to saturation as the ``memory loss'' regime.} Surprisingly, for $R\T\ll1$ we find that the formula (\ref{decay}) applies
for all values of $d$ and $\mu/T$! Similar to the case of the strip, the fact that the saturation is continuous is just a consequence of
the fact that for $R\T\ll1$, $z_*$ is \emph{uniquely} determined from the AdS embedding, and this is true regardless of the shape of the entangling surface.
The curious feature here is the increasing value of $\gamma$ with respect to the number of dimensions $d$, e.g. the second
derivative of $\Delta S_A(t)$ becomes continuous for $d\geq4$ and so on. This behavior can already be observed  from the plots in Figure \ref{EEVaidya2} $(b)$.

Another feature of our result concerns to the saturation time $t_{\text{sat}}$ itself. At the leading order of approximation, we find that $t_{\text{sat}}=z_*$ is independent of the temperature $T$ and chemical potential $\mu$. This is indeed expected because these results have been derived with the zeroth order embedding, which does not contain information about the state. However, as we will show below, the first correction to the saturation time is enough to verify the numerical behavior observed in \cite{Caceres:2012em,Caceres:2014pda}.

Before doing so, let us comment on the case of large subsystems. For the case of the strip the saturation is discontinuous
and the linear growth behavior (\ref{lgrowthb}) persists all the way to $t_{\text{sat}}$. In this case one finds that
\be\label{velargel}
v_E\simeq\frac{V_\Sigma}{A_\Sigma t_{\text{sat}}}=\frac{\ell}{2t_{\text{sat}}}+\mathcal{O}(\ell^0)\,.
\ee
Inverting equation (\ref{velargel}) gives the following expression for the saturation time at leading order:
\be
t_{\text{sat}}=\frac{\ell}{2v_E}+\mathcal{O}(\ell^0)\,.
\ee
The fact that $v_E$ decreases monotonically in $\mu/T$ leads always to an increase in $t_{\text{sat}}$.
In order to study its explicit dependence with respect to $\chi=\mu/T$ it is convenient to define
\be\label{t0def}
t^{(0)}_{\text{sat}}=\lim_{\chi\to0}t_{\text{sat}}(\chi)\,,
\ee
and normalize the result for $t_{\text{sat}}$ in units of $t^{(0)}_{\text{sat}}$ \cite{Caceres:2012em,Caceres:2014pda}. Let us consider the small $\mu/T$ limit. In this case we find that
\be\label{sattimeres}
\frac{t_{\text{sat}}}{t^{(0)}_{\text{sat}}}=1+\sigma(d)\left(\frac{\mu}{T}\right)^2+\mathcal{O}\left(\frac{\mu}{T}\right)^4\,,
\ee
where
\be
\sigma(d)=\frac{d (d-2)}{16 \pi ^2}\left[\left(\frac{d-2}{2(d-1)}\right)^{\frac{2}{d}-1}-1\right]>0\,.
\ee
So, the saturation time increases with increasing $\mu/T$, as expected. For the case of the ball (whenever the saturation is continuous) it is found that \cite{Liu:2013iza,Liu:2013qca}
\be
t_{\text{sat}}=\frac{R}{c_E}-\frac{d-2}{4\pi T}\log R +\mathcal{O}(R^0)\,,\qquad\qquad c_E=\sqrt{\frac{2\pi z_H T}{d-1}}\,.
\ee
At leading order we find a similar expression as in (\ref{sattimeres}) (with a subleading term of order $\mathcal{O}(\log R/R)\to0$) where in this case
\be\label{sattimeres2}
\sigma(d)=\frac{d(d-2)^2}{32(d-1)\pi^2}>0\,.
\ee
Again, the saturation time is found to increase with increasing chemical potential.

Let us now go back to the case of small subsystems. In the thin shell approximation, the saturation time $t_{\text{sat}}$ is given by
the time at which the vacuum extremal surface grazes the shell at $v=0$ (see Figure \ref{figthinshell}). This observation
is intuitive: for $t>t_{\text{sat}}$ the whole extremal surface lies entirely in the portion
of the geometry described by an AdS-RN black hole and, therefore, the entanglement entropy
has reached equilibrium. At the leading order in $\ell\T$ (or $R\T$) we have that $v=t-z$ so $v=0$ implies $t=z$. This is the origin of the $\theta(t-z)$ function appearing in (\ref{deltaSon}) and (\ref{deltaSon2}). The integrals are then evaluated from 0 to $z_*$ so at the end of the computation one naturally obtains $t_{\text{sat}}=z_*$, independent of $T$ or $\mu$. There are two corrections that have to be taken into account at the next order. One one hand, the translation between the Eddington-Finkelstein coordinate $v$, the boundary time $t$ and $z$ receives corrections of order $\mathcal{O}(z^{d+1})$. These corrections can be directly computed from (\ref{efcoords}). On the other hand, $z_*$ as a function of $\ell$ (or $R$) is modified as one consider corrections to the embedding above pure AdS. The full computation is explicitly carried out in Appendix \ref{correctiontsat}. For the strip, the final result reads:\footnote{In Appendix \ref{correctiontsat} we discuss some subtleties in the computation for case of the ball.}
\be\label{sattimeresult}
\frac{t_{\text{sat}}}{t^{(0)}_{\text{sat}}}=1-\left(\kappa(d)(T\ell)^d+\mathcal{O}\left(T\ell\right)^{2(d-1)}\right)\left(\frac{\mu}{T}\right)^2+\mathcal{O}\left(\frac{\mu}{T}\right)^4\,,
\ee
where
\be\label{kapadef}
\kappa(d)=\frac{(d-2)2^{d-5}\pi^{(d-4)/2}\Gamma[\frac{1}{2 (d-1)}]^d\left(\Gamma [\frac{1}{2 (d-1)}] \Gamma [\frac{d}{d-1}]-2 \Gamma [\frac{d+1}{2 (d-1)}] \Gamma [\frac{d}{2 (d-1)}]\right)}{(d+1)d^{d-2}\Gamma[\frac{d}{2 (d-1)}]^{d+1} \Gamma [\frac{d+1}{2 (d-1)}]}>0\,.
\ee
Together with equation (\ref{sattimeres}), this result confirms the numerical findings of \cite{Caceres:2012em,Caceres:2014pda}, namely that for small regions and small values of $\mu/T$ the saturation time decreases as we increase $\mu/T$ while, for large intervals, the saturation is delayed as we increase $\mu/T$.

\section{Observations for entangling surfaces of arbitrary size\label{secarb}}

\subsection{Universality of the quadratic growth regime \label{earlyuniversal}}

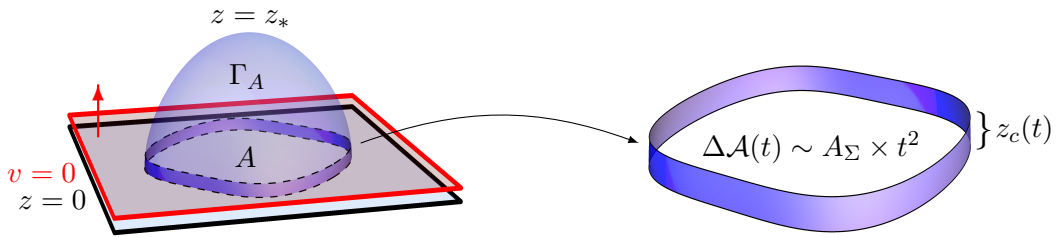
\begin{figure}[t!]
\begin{center}
\hspace{-1.5cm}\includegraphics[angle=0,width=0.55\textwidth,trim = 0 0.5cm 0 0]{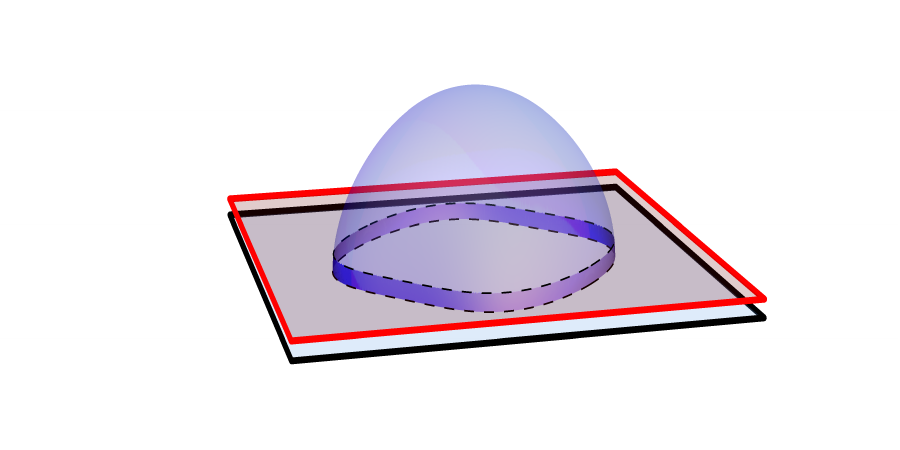}\hspace{0.3cm}
\includegraphics[angle=0,width=0.35\textwidth]{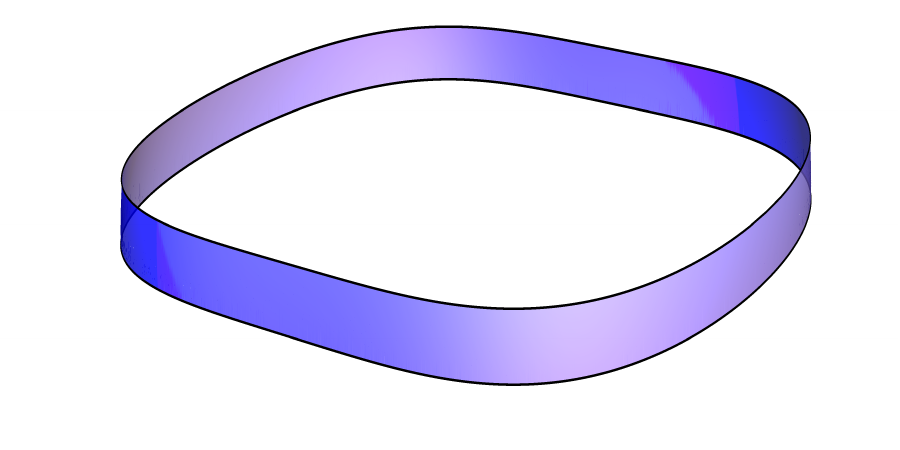}
\begin{picture}(0,0)
\put(-380,30){{\small {\color{red}$v=0$}}}
\put(-376,20){{\small $z=0$}}
\put(-304,90){{\small $z=z_*$}}
\put(-297,67){{\small $\Gamma_A$}}
\put(-295,36){{\small $A$}}
\put(-120,40){{\small $\Delta\mathcal{A}(t)\sim A_\Sigma\times t^2$}}
\put(-10,47){{\small $z_c(t)$}}
\put(-18,46){{\large $\}$}}
\thicklines
\put(-346,47){{\color{red}\vector(0,20){20}}}
\end{picture}
\end{center}
\vspace{-0.7cm}
\caption{\small Computation of the entanglement entropy growth at early times. All extremal surfaces intersect the boundary of AdS at right angle (regardless of the size or shape of $\Sigma$), so the leading contribution is simply $A_\Sigma\times z_c(t)$ (where $z_c(t)$ is the position of the shell at time $t$) times a conformal factor that may only depend on $z_c(t)$. The leading correction to the geometry near the boundary comes with a factor of $z_H^{-d}\sim\T^{d}\sim\mathcal{E}$. Via dimensional analysis, this fixes the initial time dependence to be $t^2$.}
\label{figthinshellearly}
\begin{center}
\vspace{-5.9cm}
\begin{tikzpicture}
        \draw (-11.05,100) node[circle, inner sep=0.8pt] (C) {};
        \draw (-7.3,100) node[circle, inner sep=0.8pt] (D) {};
        \hspace{-0.25cm}
        \draw[->] (C) to [bend left=20] (D);
\end{tikzpicture}
\end{center}
\vspace{4.4cm}
\end{figure}

The fact that the initial growth regime (\ref{quda}) shows no dependence with the size or shape of the entangling region suggests that this behavior may be universal. Via dimensional analysis, we can infer that in a quadratic growth regime, the coefficient of the $t^2$ must be given by the area of $\Sigma$, $A_\Sigma$, times a dimensionless coefficient that may depend on the shape of $\Sigma$. It is easy to see that this coefficient is indeed independent of $\Sigma$. For $t\ll t_{\text{sat}}$ the shell is very close to the boundary so the relevant contribution comes from the near boundary portion of the geometry. Since, all extremal surfaces intersect the boundary of AdS at right angle (regardless of the shape of $\Sigma$), the leading contribution at early times for the change in $\Delta\A(t)$ is simply $A_\Sigma\times z_c(t)$ (where $z_c(t)$ is the position of the shell at time $t$) times a conformal factor that may only depend on $z_c(t)$. This proves that $A_\Sigma$ is the only dependence of $\Sigma$ in the early time regime. In addition, since the leading correction from AdS near the boundary has a factor of $z_H^{-d}\sim\T^{d}\sim\mathcal{E}$ then, by dimensional analysis, it follows that the time dependence in this regime must be $t^2$ (see Figure \ref{figthinshellearly}).

A direct calculation of the early-time growth for a general $\Sigma$ was done in \cite{Liu:2013qca} and the final formula can be written in terms of the energy density as follows:
\be
\Delta S_A(t) =\frac{\pi}{d-1}\mathcal{E} A_\Sigma t^2+ \cdots \,.
\ee
Indeed, we can verify that with our formula for the energy density (\ref{energydens}) and the high and low effective temperature expansions (\ref{zhthermal})-(\ref{zhquantum}) we can recover the appropriate early time growth for near-thermal and near-extremal quenches (\ref{igthermal})-(\ref{igextremal}).


\subsection{Bound on the saturation time\label{saturationsec}}

In this section we will provide a bound on $t_{\text{sat}}$ in different corners of the space of parameters, specializing to the case of the strip.
In order to obtain the bound we compute the time $t^*_{\text{sat}}$ at which there is a solution which lies fully in the back hole region. If the saturation is continuous then $t^*_{\text{sat}}=t_{\text{sat}}$ but for discontinuous saturation one finds that $t^*_{\text{sat}}\leq t_{\text{sat}}$ \cite{Liu:2013iza,Liu:2013qca}, so it provides a lower bound. From equation (\ref{efcoords}) it follows that
\be\label{tsatint}
t^*_{\text{sat}}=\int_0^{z_*}\frac{dz}{f(z)}\,.
\ee
We also need the function $z_*(\ell)$. Fortunately, at $t=t^*_\text{sat}$ the entire surface
lies entirely in a static AdS-RN background, so the problem is time-independent. In order to obtain
$z_*(\ell)$ we use the fact that for the strip we have a conservation equation (since the
area functional does not depend explicitly on $x$):
\be
x'(z)=\pm\frac{1}{\sqrt{f(z)[(z_*/z)^{2(d-1)}-1]}}\,.
\ee
Therefore, from the boundary condition (\ref{bc}) it follows that\footnote{For the ball we do not have a conservation equation, so we cannot use the same methodology.}
\be\label{lint}
\ell=2\int_0^{z_*}\frac{dz}{\sqrt{f(z)[(z_*/z)^{2(d-1)}-1]}}\,.
\ee
This last equation must be solved and inverted to obtain $z_*(\ell)$. Now, following \cite{Fischler:2012ca} we can formally write (\ref{tsatint}) as double sum:
\be\label{tsatseries}
t^*_{\text{sat}}=z_*\sum _{n=0}^{\infty }\sum _{k=0}^{n} \frac{(-1)^k\varepsilon^{n-k} (\varepsilon-1)^k \Gamma[n+1]}{(1+n d+(d-2) k)\Gamma[k+1]\Gamma[n-k+1]}\left(\frac{z_*}{z_H}\right)^{n d+ k(d-2) }\,.
\ee
Similarly from \cite{Kundu:2016dyk}, we can write (\ref{lint}) as
\begin{align}
\ell=\frac{z_*}{d-1}\sum_{n=0}^\infty \sum_{k=0}^n \frac{\varepsilon^{n-k} (\varepsilon-1)^k\Gamma\left[\frac{2n+1}{2}\right]\Gamma \left[\frac{d (n+k+1)-2k}{2 (d-1)}\right]}{
\Gamma[1+n-k]\Gamma[k+1]\Gamma \left[\frac{d (n+k+2)-2k-1}{2 (d-1)}\right]}  \left(\frac{z_*}{z_H}\right)^{n d +k(d-2)} \ .\label{eerc2}
\end{align}
In the following we will use these expansion to compute the saturation time in various regimes.

\subsection*{Small subsystems}

For $\ell\T\ll1$ we expect continuous saturation. In this case
\be
t^*_{\text{sat}}=t_{\text{sat}}=z_*\left[1+\frac{\varepsilon}{d+1} \left(\frac{z_*}{z_H}\right)^d+\O\left(\frac{z_*}{z_H}\right)^{2(d-1)}\right]\,,
\ee
while
\be
\ell=\frac{2 \sqrt{\pi }\Gamma[\frac{d}{2(d-1)}]z_*}{\Gamma[\frac{1}{2(d-1)}]}\left[1+\frac{\varepsilon\Gamma[\frac{d}{d-1}] \Gamma [\frac{1}{2 (d-1)}]}{2(d+1)\Gamma [\frac{d+1}{2(d-1)}] \Gamma[\frac{d}{2(d-1)}]}\left(\frac{z_*}{z_H}\right)^d+\O\left(\frac{z_*}{z_H}\right)^{2(d-1)}\right].
\ee
This last equation can be inverted perturbatively to obtain:
\be\label{zstcorrected}
z_*=\frac{\Gamma[\frac{1}{2 (d-1)}]\ell}{2\sqrt{\pi} \Gamma[\frac{d}{2 (d-1)}]}\left[1-\frac{\varepsilon\Gamma[\frac{d}{d-1}] \Gamma[\frac{1}{2 (d-1)}]^{d+1}}{2^{d+1}(d+1)\pi^{d/2} \Gamma[\frac{d+1}{2(d-1)}] \Gamma[\frac{d}{2 (d-1)}]^{d+1}}\left(\frac{\ell}{z_H}\right)^{d}+\O\left(\frac{\ell}{z_H}\right)^{2(d-1)}\right].
\ee
Therefore, at the leading order we obtain,
\be\label{sattime22}
t_{\text{sat}}=\frac{\ell~\Gamma[\frac{1}{2(d-1)}]}{2 \sqrt{\pi}\Gamma[\frac{d}{2(d-1)}]}\left[1+\varepsilon\tau_1(d)(\l\T)^d+\O(\l\T)^{2(d-1)}\right]\ ,
\ee
where, $\tau_1(d)$ is the following numerical factor
\be
\tau_1(d)=\frac{2^d \pi ^{d/2} \Gamma[\frac{1}{2 (d-1)}]^d  }{d^d (d+1)\Gamma [\frac{d}{2 (d-1)}]^{d}}\left(1-\frac{2^{\frac{1}{d-1}-1} \Gamma[\frac{1}{2 (d-1)}] \Gamma [\frac{2 d-1}{2 (d-1)}]}{\sqrt{\pi } \Gamma [\frac{d+1}{2 (d-1)}]}\right)<0\ .
\ee

Let us now consider different regimes of the above saturation time. In the limit $\mu/T\ll1$ we obtain
\be
t_{\text{sat}}=\frac{\ell\Gamma[\frac{1}{2(d-1)}]}{2 \sqrt{\pi}\Gamma[\frac{d}{2(d-1)}]}\left[1+\tau_1(d) (\l T)^d\left(1+\frac{d^2(d-2)}{16\pi^2}\left(\frac{\mu}{T}\right)^2\right)+\cdots\right]\,,
\ee
which can be rewritten as
\be
t_{\text{sat}}=t_{\text{sat}}^{(0)}\left[1-\kappa(d) (\l T)^d\left(\frac{\mu}{T}\right)^2+\cdots\right]\,,\qquad t_{\text{sat}}^{(0)}\equiv\frac{\ell\Gamma[\frac{1}{2(d-1)}]}{2 \sqrt{\pi}\Gamma[\frac{d}{2(d-1)}]}\,,
\ee
The constant $\kappa(d)$ is given in (\ref{kapadef}) and is positive. Therefore, the saturation time decreases with the increase of chemical potential.

A similar result can also be obtained in the limit $\mu/T\gg1$. From (\ref{sattime22}) it follows that, for $T=0$:
\be
\tilde{t}_{\text{sat}}^{(0)}=\frac{\ell\Gamma[\frac{1}{2(d-1)}]}{2 \sqrt{\pi}\Gamma[\frac{d}{2(d-1)}]}\left[1+\frac{2(d-1)\tau_1(d)}{(d-2)} \left(\frac{d^{1/2}(d-2)}{4 \pi (d-1)^{1/2}}\right)^d(\mu \l)^d\right]\ .
\ee
Now, for $\mu/T\gg1$ we obtain
\be
t_{\text{sat}}=\tilde{t}_{\text{sat}}^{(0)}\left[1-\frac{\tau_1(d)}{2(d-1)}\left(\frac{d^{1/2}(d-2)}{4 \pi (d-1)^{1/2}}\right)^{d-1} (\mu \l)^d\left(\frac{T}{\mu}\right)+\cdots\right]\ .
\ee
Since $\tau_1(d)$ is negative, the saturation time increases with the increase of temperature.

\subsection*{Large subsystems}

The limit $\ell\T\gg1$ corresponds to $z_*\rightarrow z_H$. In this case the saturation can be discontinuous in some cases so $t^*_{\text{sat}}$ provides a lower bound for the actual saturation time $t_{\text{sat}}$ \cite{Liu:2013iza,Liu:2013qca}. It is easy to check that in this limit both $\l$ and $t^*_{\text{sat}}$ diverge. However, we can define a combination of  $\l$ and $t^*_{\text{sat}}$ which is finite as we let $z_*\rightarrow z_H$:
\be\label{large}
t^*_{\text{sat}}-\l\sqrt{\frac{(d-1)}{2(d-2)\delta}}=\int_0^{z_*}dz \left[\frac{1}{f(z)}-\sqrt{\frac{2(d-1)}{(d-2)\delta}}\frac{1}{\sqrt{f(z)[(z_*/z)^{2(d-1)}-1]}}\right]\,,
\ee
where
\be
\delta=\frac{2(d-1)}{(d-2)}-\varepsilon\,.
\ee
Before we proceed, a few comments are in order: the right hand side of (\ref{large}) is finite in the limit $z_*\rightarrow z_H$ and hence we can write
\be\label{result1}
t^*_{\text{sat}}=\l\left(\sqrt{\frac{(d-1)}{2(d-2)\delta}}+\frac{ \tau_2(d, \delta)\ d}{4\pi \T\l}\right)\ ,
\ee
where,
\be
\tau_2(d, \delta)=\int_0^{1}dx \left[\frac{1}{f(x z_H)}-\sqrt{\frac{2(d-1)}{(d-2)\delta}}\frac{1}{\sqrt{f(x z_H)[(1/x)^{2(d-1)}-1]}}\right]\ .
\ee
Secondly, the limit $\delta\to0$ (or $T=0$) appears to be singular. Indeed, in this case $t^*_{\text{sat}}$ is no longer linear in $\l$
and is expected to grow at a faster rate \cite{Albash:2010mv}; we will consider this case separately. Before doing so, let us consider the case $\mu/T\ll1$. For $\mu=0$ we have
\be
t_{\text{sat}}^{*(0)}=\sqrt{\frac{d-1}{2d}}\l\left[1+\O\left(\frac{1}{\l T}\right) \right]\ .
\ee
Now, for $\mu/T\ll1$ we obtain
\be\label{tsatcont}
t^*_{\text{sat}}=t_{\text{sat}}^{*(0)}\left[1+\frac{d(d-2)^2  }{32 \pi ^2 (d-1) }\left(\frac{\mu}{T} \right)^2+\O\left(\frac{1}{\l T}\right)+\O\left(\frac{\mu}{T}\right)^4\right]\,,
\ee
which increases with the chemical potential. Notice that (\ref{tsatcont}) is the result that we obtained for the case of the ball (\ref{sattimeres2}). This suggests that (\ref{tsatcont}) gives the actual saturation time for all shapes, provided that the saturation is continuous.\footnote{Let us assume that $t_{\text{sat}}^{*(0)}$ is known for a specific shape. Since the first correction in $\mu/T$ is independent of $\ell$, the result at this order should be independent of the precise definition of $\ell$, i.e. it can be taken as a characteristic length scale of the subsystem.} Also note that equation (\ref{tsatcont}) is different from the actual saturation time (\ref{sattimeres}), which tells us that the saturation is discontinuous for strips of length $\ell\gg1/\T$.

Finally, let us consider the $T=0$ case. Assuming that $z_*=z_H(1-\epsilon)$ with $\epsilon\ll1$, it is easy to show that in this case
\be
t^*_{\text{sat}}=\frac{d}{4\pi\T}\left[\frac{1}{d(d-1)  \epsilon }+\frac{(3 d-5) \log \epsilon+3}{3 (1-d) d}+\tau_3(d)\right]\ ,
\ee
where $\tau_3(d)$ is the finite integral
\be
\tau_3(d)=\int_0^{1}dx\left[\frac{1}{f(x z_H)}+\frac{3 d (x-1)-5 x+2}{3 (d-1) d (x-1)^2}\right]_{\delta=0}\,.
\ee
Similarly, in this limit one can also show
\be
\l=\frac{2d}{4\pi\T}\left[\frac{\sqrt{2}}{(d-1) \sqrt{d} \sqrt{\epsilon }}+\frac{\sqrt{2}}{(1-d) \sqrt{d}}+\tau_4(d)\right]\ ,
\ee
where $\tau_4(d)$ is another finite integral
\be
\tau_4(d)=\int_0^{1}dx \left[\frac{1}{\sqrt{f(x z_H)[(1/x)^{2(d-1)}-1]}}-\frac{1}{\sqrt{2 d} (d-1) (1-x)^{3/2}}\right]_{\delta=0}\ .
\ee
Therefore, at the leading order
\be
\epsilon=\frac{d}{2 \pi ^2 (d-1)^2 \l^2\T^2}
\ee
and hence
\be\label{tsat0}
t^*_{\text{sat}}=\frac{\pi  (d-1)  \T \l^2}{2 d}=\frac{(d-2) \sqrt{\pi(d-1) d}\ \mu \l^2}{8 d}\ .
\ee
Therefore, in this limit $t^*_{\text{sat}}$ increases with the chemical potential. Our result (\ref{tsat0}) is also
consistent with the numerical results of \cite{Albash:2010mv} regarding the fast growth of the saturation time with respect to the length $\ell$.

\subsection{Bound on the average velocity from bulk causality\label{boundsec}}

Let us now discuss the average velocity in more generality. In Section \ref{quasilinear} we showed that $v^{\text{avg}}_E$
is a better quantity to consider when comparing results between entangling regions of different sizes. We further conjectured that, even though $v_E^{\text{max}}$ can exceed the speed of light, $v^{\text{avg}}_E$ should be constrained by causality. In the limit of large regions $v_E=v^{\text{avg}}_E$ so the bound derived in \cite{Casini:2015zua,Hartman:2015apr} is directly applicable. For small regions the bound seems to be satisfied at least for the strip and the ball so it is very likely that
\be\label{boundavg}
v^{\text{avg}}_E\leq1
\ee
holds more generally. Here, we argue that such a bound is a direct consequence of bulk causality. To see this, consider the formula for the average velocity:
\be
v^{\text{avg}}_E=\frac{V_A}{A_\Sigma t_\text{sat}}\,.
\ee
For the case of the strip, the ratio $V_A/A_\Sigma=\ell/2=t_{\text{light}}$ is equal to the light-crossing time from $\Sigma$ to the interior of the region $A$. So, in order to decide if (\ref{boundavg}) is satisfied or not we have to compute $t_\text{sat}$ and compare it with $t_{\text{light}}$. Quite generally, we find that
\be\label{tsatineq}
t_{\text{sat}}\geq t^*_{\text{sat}}=\int_0^{z_*}\frac{dz}{f(z)}\geq\int_0^{z_*}dz=z_*\geq t_{\text{light}}\,.
\ee
The first part of this equation comes from the definition of $t^*_{\text{sat}}$ (\ref{tsatint}) which gives us a bound on the saturation time $t_{\text{sat}}$. At $t=t^*_{\text{sat}}$ there is an extremal surface which lies fully in the back hole region, for which $v\geq0$.
The shell is located at $v=0$ and is moving at the speed of light; however, due to the redshift factor $f(z)\leq1$, we obtain that $t^*_{\text{sat}}\geq z_*$. The last part of equation (\ref{tsatineq}) comes from a comparison of the extremal surface $\Gamma_A$ and the causal wedge $\Xi_A$ associated to $A$ \cite{Hubeny:2012wa}. In this paper it was found that the causal wedge $\Xi_A$ always lies closer to the boundary than the extremal surface $\Gamma_A$, so $z_*\geq z^{\Xi}_*\geq t_{\text{light}}$. Putting everything together, then, we conclude that for the strip $v^{\text{avg}}_E\leq1$. For other geometries (\ref{tsatineq}) is still true but the ratio $V_A/A_\Sigma$ may vary. For finite subsystem, the volume-to-area ratio is maximized for the case of the ball, for which $V_A/A_\Sigma=R/(d-1)=t_{\text{light}}/(d-1)$.\footnote{This is a consequence of the isoperimetric inequality, see e.g. \cite{Federedbook}.} Therefore, $v^{\text{avg}}_E\leq1$ still holds. For convex strips the volume-to-area ratio is maximized for the case of the rectangular strip, which we already consider. Finally, for concave strips the ratio can be higher but these are considered as large subsystems so, again, $v^{\text{avg}}_E\leq1$. This conclude our proof of (\ref{boundavg}).

\section{Conclusions\label{concsec}}

In this paper we developed new analytical tools to study the thermalization of entanglement entropy after
a global quench in the context of the AdS/CFT correspondence. We focused on the limit of small
subsystems, for which no previous technique was available in the literature, and found some surprising
results.

In Section \ref{2dsec} we began our investigation by exploring the known analytical results for $(1+1)-$dimensional holographic CFTs,
focusing on the different regimes of interest. We pointed out that the conjectured bound on the maximum rate of growth for the entanglement
entropy only holds in the strict limit of large intervals, but is violated otherwise. In particular, we found that $\text{max}[\mathfrak{R}(t)]\to1$
as we let $\mathfrak{l}\to\infty$ but it generally exceeds the speed of light for intervals of finite size. We also observed that the linear growth regime is smoothed out as we reduce the size of the system, suggesting that the interpretation in terms of a ``entanglement tsunami'' is no longer valid. In Section \ref{sectemp} we introduced holographic models of global quenches in higher dimensions: CFT states dual to a collapsing AdS-RN-Vaidya geometry. We specialized to the thin shell regime, which is valid for instantaneous quenches. In Section \ref{secevol} we computed perturbatively
the evolution of entanglement entropy after the quench focusing on two different entangling surfaces: the strip and the ball. At this point it became clear that: $i)$ the violation of the inequality $\text{max}[\mathfrak{R}(t)]\leq1$ is only present in $(1+1)$ dimensions $ii)$ the initial and final stages of the evolution are always smooth and $iii)$ the evolution in the intermediate regime depends on the shape of the entangling region but is insensitive to the final state of the quench.

In Section \ref{secregimes} we studied more in detail our results for the strip and the ball in different regimes of the thermalization process. For the early time regime, the evolution turned out to be independent of the entangling region and in agreement with the results for large subsystems. This observation led us to conjecture that the evolution in this regime is universal and completely fixed by symmetries. In the intermediate regime
we found a non-universal quasi-linear growth regime with a maximum rate of growth $v_E^{\text{max}}$ that depends on the shape of the entangling region.
The maximum rate is found to be higher for small intervals in $d\leq4$ (strip) or $d\leq3$ (ball) as long as $\mu/T\ll1$, but is lower in higher dimensions. For $\mu/T\gg1$ the maximum rate is always higher for small intervals. We pointed out that the average velocity $v_E^{\text{avg}}$ is a better parameter if we are to compare results for entangling regions of different sizes. We found that, in average, the UV degrees of freedom equilibrate at a slower rate when the evolution is governed by thermal fluctuations ($\mu/T\ll1$) but at a faster rate if the evolution is driven by quantum fluctuations ($\mu/T\gg1$). This conclusion is more robust than the one for $v_E^{\text{max}}$ because it is independent of the number of dimensions and the shape of the entangling region. Moreover, as we proved in the last section, $v_E^{\text{avg}}$ is actually constrained by causality. The approach to saturation is found to be always continuous and is characterized by a nontrivial scaling exponent that depends on the number of dimensions and the shape of the entangling region. We explain this by arguing that, at the leading order, $z_*$ is uniquely determined by the embedding pure AdS. However, for large subsystems $z_*$ may be multi-valued near the saturation time, leading to a discontinuous behavior. We also computed the leading correction to $t_{\text{sat}}$ and confirmed the non-monotonicity with respect to $\mu/T$ observed numerically in \cite{Caceres:2012em,Caceres:2014pda}.

In Section \ref{secarb} we made some general remarks about entangling surfaces of arbitrary size. We started by giving a simple argument to explain the universality of the initial quadratic growth regime. The physical picture is the following: all extremal surfaces intersect the boundary of AdS at right angle (regardless of the size or shape of the entangling region), so the leading contribution at early times is simply $A_\Sigma\times z_c(t)$ (where $z_c(t)$ is the position of the shell at
time $t$) times a conformal factor that may only depend on $z_c(t)$. The leading correction to the
geometry near the boundary comes with a factor of $z_H^d\sim\mathcal{E}$, which in turn fixes the initial time dependence to be $t^2$.
Later in the same section, we gave a simple recipe for computing a bound on the saturation time in different regimes of interest. Using this method, we were able to study the saturation time in various limits and to corroborate its non-trivial dependence with respect to the chemical potential. At the end of the section we provided a proof for a bound on $v_E^{\text{avg}}$ based on bulk causality. We believe that this bound should hold more generally, as long as the theory is relativistically invariant.

There are various open questions and a number of possibilities for the extension of this work.
The most urgent one is to investigate possible bounds on $v_E^{\text{max}}$ and $v_E^{\text{avg}}$ from the
field theory perspective, i.e. generalize the analysis of \cite{Casini:2015zua,Hartman:2015apr}
for entangling regions of arbitrary size. In particular, the interacting models of \cite{Casini:2015zua}
seem a good staring point for this investigation. Another interesting possibility is to consider the case of
$(1+1)-$dimensional CFTs at large central charge, where the conformal block expansion has proved to be an
efficient tool \cite{Asplund:2015eha}. Moving to the realm of holography, we can consider gravity duals
of theories with different symmetries. Of particular interest are the non-relativistic theories with
Lifshitz scaling and/or hyperscaling violation \cite{Alishahiha:2014cwa,Fonda:2014ula}, which have recently gained
attention in the context of AdS/CMT. We can also consider CFTs on a sphere; interestingly, charged solutions in global AdS have been shown to exhibit a very rich entanglement phase structure \cite{Johnson:2013dka,Caceres:2015vsa}.
Finally, we can use the techniques developed here to study the thermalization of
other field theory observables after a global quench, e.g. two-point functions \cite{Aparicio:2011zy}, Wilson loops \cite{Balasubramanian:2011ur}, and other entanglement related quantities such as mutual information \cite{Allais:2011ys,Alishahiha:2014jxa,Tanhayi:2015cax}, causal holographic information \cite{Hubeny:2013hz} and holographic complexity \cite{Alishahiha:2015rta}.
We hope to return to some of these problems in the near future \cite{progess}.

\section*{Acknowledgements}
It is a pleasure to thank Mohsen Alishahiha, Elena Caceres, Jan de Boer, Ben Freivogel, Arnab Kundu, Sagar Lokhande and Gerben Oling for discussions and comments on the manuscript. SK is supported by the NSF grant PHY-1316222. JFP is supported by the Foundation for Fundamental Research on Matter (FOM) which is part of the Netherlands Organization for Scientific Research (NWO). JFP would also like to thank the Departament de F\'isica Fonamental at Universitat de Barcelona for the warm hospitality during the final stages of this work.

\appendix

\section{Perturbative computation at next-to-leading order\label{nexttoleading}}

Based on the expansion given in (\ref{arealambdaonshell}) we expect that the first correction due to the corrected embedding will appear at order $\mathcal{O}(\lambda^2)$; this is indeed expected since this correction arises from the combination of both $\phi^{(1)}$ and $\mathcal{L}^{(1)}$, which are of order $\mathcal{O}(\lambda)$. However, due to the particular form of the metric (\ref{metricexpansion}) we can see that the second correction to the functional $\mathcal{L}$ is actually of order $\mathcal{O}(\lambda^{2-2/d})\gg \mathcal{O}(\lambda^2)$ for any finite $d$.  Therefore, at this order of approximation the correction to the embedding is still negligible and we can still use the solution for pure AdS!

The computations are very similar to the ones presented in Section \ref{pertcomp}, so we will only sketch the main few steps, specializing to the two geometries in consideration, the strip and the ball.

\subsection*{The strip}

Expanding the area functional (\ref{larangian}) to the next-to-leading order we get,
\be
\mathcal{L}^{(2-2/d)}=-\frac{(\varepsilon-1)A_\Sigma}{2z_H^{2(d-1)}}\frac{z^{d-1} v'^2 \theta(v)}{\sqrt{x'^2-v'^2-2v'}}\,.
\ee
Evaluating it on shell, this yields the following contribution to the entanglement entropy:
\be\label{ent2}
\Delta S_A^{(2-2/d)}(t)=-\frac{(\varepsilon-1)A_\Sigma}{8G_N^{(d+1)}z_H^{2(d-1)}}\int_0^{z_*}dz\, \theta (t-z) z^{d-1} \sqrt{1-(z/z_*)^{2(d-1)}}\,.
\ee
The integral in (\ref{ent2}) is reminiscent of the one appearing in (\ref{deltaSon}) and can be evaluated in a similar way. The final result can be written as follows:
\be\label{deltaSfinal1b}
\Delta S_A^{(2-2/d)}(t)=\Delta S_{\text{eq}}^{(2-2/d)}\big\{[\theta(t)-\theta(t-t_{\text{sat}})]\tilde{\mathcal{F}}(t/t_{\text{sat}})+ \theta(t-t_{\text{sat}})\big\}\,,
\ee
where $\Delta S_{\text{eq}}^{(2-2/d)}$ is given by,
\be
\Delta S_{\text{eq}}^{(2-2/d)}=-\frac{(d-1)\sqrt{\pi}\Gamma[\frac{3d-2}{2(d-1)}]z_*^d A_\Sigma(\varepsilon-1)}{8d(2d-1)\Gamma[\frac{2d-1}{2(d-1)}]z_H^{2(d-1)}G_N^{(d+1)}}\,,
\ee
and
\be\label{deff2un}
\tilde{\mathcal{F}}(x)=\frac{d\,\Gamma[\frac{2d-1}{2(d-1)}]x^d}{(d-1)\sqrt{\pi}\Gamma[\frac{3d-2}{2(d-1)}]} \left[\sqrt{1-x^{2(d-1)}}+\tfrac{d-1}{d}\!\,_2F_1\left(\tfrac{1}{2},\tfrac{d}{2(d-1)},\tfrac{3d-2}{2(d-1)},x^{2(d-1)}\right)\right].
\ee

\subsection*{The ball}

Expanding the area functional (\ref{larangian2}) for the ball we get,
\be
\mathcal{L}^{(2-2/d)}=-\frac{(\varepsilon-1) A_\Sigma}{2R^{d-2}z_H^{2(d-1)}}\frac{z^{d-1}r^{d-2} v'^2 \theta(v)}{\sqrt{r'^2-v'^2-2v'}}\,.
\ee
Evaluating it on shell leads to the following contribution to the entanglement entropy:
\be\label{ent2}
\Delta S_A^{(2-2/d)}(t)=-\frac{(\varepsilon-1) A_\Sigma z_*^{d-2}}{8G_N^{(d+1)}R^{d-2}z_H^{2(d-1)}}\int_0^{z_*}dz\, \theta (t-z) z^{d-1} \left[1-\left(z/z_*\right)^2\right]^{\frac{d-1}{2}}\,.
\ee
Finally, performing the integration we obtain:
\be\label{deltaSfinal2b}
\Delta S_A^{(2-2/d)}(t)=\Delta S_{\text{eq}}^{(2-2/d)}\big\{[\theta(t)-\theta(t-t_{\text{sat}})]\tilde{\mathcal{G}}(t/t_{\text{sat}})+ \theta(t-t_{\text{sat}})\big\}\,,
\ee
where in this case
\be
\Delta S_{\text{eq}}^{(2-2/d)}=
-\frac{\sqrt{\pi }\Gamma[d]R^dA_{\Sigma}(\varepsilon -1)}{2^{d+3}\Gamma[\frac{2d+1}{2}]z_H^{2(d-1)}G_N^{(d+1)}}\,,
\ee
and
\be\label{defg2un}
\tilde{\mathcal{G}}(x)=\frac{2^d x^d \Gamma[\frac{2d+1}{2}]}{\sqrt{\pi }d \Gamma[d]} \,\!_2F_1\left(-\tfrac{d-1}{2},\tfrac{d}{2},\tfrac{d+2}{2},x^2\right)\,.
\ee

\section{First correction to the saturation time\label{correctiontsat}}

Let us start by considering equation (\ref{efcoords}). In the black hole portion of the geometry
\bea
&&v=t-\int_0^z\frac{dz'}{f(z')}=t-\int_0^zdz'\left[1+\varepsilon\left(\frac{z'}{z_H}\right)^d+\mathcal{O}\left(\frac{z'}{z_H}\right)^{2(d-1)}\right],\nonumber\\
&&v=t-z\left[1+\frac{\varepsilon}{(d+1)}\left(\frac{z}{z_H}\right)^d+\mathcal{O}\left(\frac{z}{z_H}\right)^{2(d-1)}\right].
\eea
Thus, evaluating at $v=0$ and $z=z_*$ we get
\be\label{tsatcorrected}
t_{\text{sat}}=z_*\left[1+\frac{\varepsilon}{(d+1)}\left(\frac{z_*}{z_H}\right)^d+\mathcal{O}\left(\frac{z_*}{z_H}\right)^{2(d-1)}\right]\,.
\ee
Let us now compute the corrections to $z_*$. In the following we will specialize to the two cases in consideration, namely the strip and the ball.

\subsection*{The strip}
In order to find the corrections to $z_*(\ell)$ we have to solve the equations of motion that come from (\ref{larangian}) at next-to-leading order. Fortunately, since at $t=t_{\text{sat}}$ the entire surface lies entirely in the black hole portion of the geometry we can consider solving the problem in a static AdS-RN geometry. For the strip we have a conservation equation since the lagrangian does not depend explicitly on $x$:
\be\label{fulleqn}
x'(z)=\pm\frac{(z/z_*)^{d-1}}{\sqrt{f(z)} \sqrt{1-(z/z_*)^{2(d-1)}}}\,.
\ee
The embedding is even with respect to $x\to-x$ so without loss of generality, we will consider the $(-)$ sign in (\ref{fulleqn}) (this corresponds to the $x>0$ portion of the embedding). Evidently, all the corrections over AdS come from the $f(z)$ term so we can expand all terms as in (\ref{metricexpansion}). More specifically, we consider
\be\label{metricexpansion2}
f(z)=1-\varepsilon\left(\frac{z}{z_H}\right)^d\zeta^d+\mathcal{O}(\zeta^{2(d-1)})\,,
\ee
and
\be\label{xexpansion2}
x(z)=x_{0}(z)+ x_{d}(z)\zeta^d+\mathcal{O}(\zeta^{2(d-1)})\,,
\ee
and at the end we set $\zeta\to1$. Plugging (\ref{metricexpansion2}) and (\ref{xexpansion2}) back into (\ref{fulleqn}) we get the following equations at leading and next-to-leading order:
\be
x_{0}'(z)=-\frac{(z/z_*)^{d-1}}{\sqrt{1-(z/z_*)^{2(d-1)}}}\,,
\ee
and
\be\label{eqcorrect}
x_{d}'(z)=-\frac{\varepsilon}{2}\left(\frac{z}{z_H}\right)^d\frac{(z/z_*)^{d-1}}{\sqrt{1-(z/z_*)^{2(d-1)}}}\,,
\ee
respectively. The solution for $x_{0}(z)$ part is given in (\ref{embAdS}), namely
\be\label{embAdS}
x_{0}(z)=\frac{\ell}{2}-\frac{z_*}{d}\left(\frac{z}{z_*}\right)^d\,\!_2F_1\left[\frac{1}{2},\frac{d}{2(d-1)},\frac{3d-2}{2(d-1)},\left(\frac{z}{z_*}\right)^{2(d-1)}\right]\,.
\ee
For now we do not assume any relation between $\ell$ and $z_*$. Since $x_0(z)$ already satisfy the boundary condition (\ref{bc}), we have to solve (\ref{eqcorrect}) subject to the constraint $x_d(0)=0$. The solution is the following:
\be
x_d(z)=\frac{\varepsilon z_*^{d-1}z^2}{2(d+1)z_H^d}\left[\sqrt{1-\left(\frac{z}{z_*}\right)^{2(d-1)}}-\,\!_2F_1\left(\frac{1}{2},\frac{1}{d-1},\frac{d}{d-1},\left(\frac{z}{z_*}\right)^{2(d-1)}\right)\right]\,.
\ee
Next, imposing that $x(z_*)=0$ we get the following relation between $\ell$ and $z_*$:
\be
\ell=\frac{2 \sqrt{\pi }\Gamma[\frac{d}{2(d-1)}]z_*}{\Gamma[\frac{1}{2(d-1)}]}\left[1+\frac{\varepsilon\Gamma[\frac{d}{d-1}] \Gamma [\frac{1}{2 (d-1)}]}{2(d+1)\Gamma [\frac{d+1}{2(d-1)}] \Gamma[\frac{d}{2(d-1)}]}\left(\frac{z_*}{z_H}\right)^d+\mathcal{O}\left(\frac{z_*}{z_H}\right)^{2(d-1)}\right].
\ee
This equation can be inverted perturbatively to obtain:
\be\label{zstcorrected}
z_*=\frac{\Gamma[\frac{1}{2 (d-1)}]\ell}{2\sqrt{\pi} \Gamma[\frac{d}{2 (d-1)}]}\left[1-\frac{\varepsilon \Gamma[\frac{d}{d-1}] \Gamma[\frac{1}{2 (d-1)}]^{d+1}}{2^{d+1}(d+1)\pi^{d/2} \Gamma[\frac{d+1}{2(d-1)}] \Gamma[\frac{d}{2 (d-1)}]^{d+1}}\left(\frac{\ell}{z_H}\right)^{d}+\mathcal{O}\left(\frac{\ell}{z_H}\right)^{2(d-1)}\right].
\ee
Plugging (\ref{zstcorrected}) into (\ref{tsatcorrected}) we can easily get the first correction to $t_{\text{sat}}$. After some algebra, we finally arrive to (\ref{sattimeresult}).

\subsection*{The ball}
We can repeat the same steps for the case of the ball in order to get the corrections to $z_*(R)$. However, in this case we do not have a conservation law so we have to solve a second order differential equation. Again, we use (\ref{metricexpansion2}) and expand the embedding as
\be\label{sphexpa}
r(z)=r_{0}(z)+ r_{d}(z)\zeta^d+\mathcal{O}(\zeta^{2(d-1)})\,.
\ee
At the end we restore $\zeta\to1$. At the leading order, the equation of motion is
\be
r_{0}''(z)-\frac{(d-1)}{z}r_{0}'(z)^3-\frac{(d-2)}{r_{0}(z)}r_{0}'(z)^2-\frac{2(d-1)}{2 z}r_{0}'(z)-\frac{(d-2)}{r_{0}(z)}=0\,,
\ee
and the solution is the standard spherical cap (\ref{sphericalsol}),
\be
r_{0}(z)=\sqrt{z_*^2-z^2}\,.
\ee
This solution satisfies the IR boundary condition, $r_{0}(z_*)=0$. For now we do not assume any relation between $z_*$ and $R$. The equation of motion for the second term is:
\be
r_{d}''(z)-\frac{(d-1)R^2+2z^2}{z(R^2-z^2)}r_{d}'(z)+\frac{(d-2)R^2}{(R^2-z^2)^2}r_{d}(z)=\frac{\varepsilon z^d ((d-4)R^2+(d+2)z^2)}{2 z_H^d(R^2-z^2)^{3/2}}\,,
\ee
which has to solved subject to the constraint $r_d(z_*)=0$. The solution is the following:
\be
r_{d}(z)=\frac{\varepsilon }{z_H^d}\left(\frac{2z_*^{d+2}-z^d(z_*^2+z^2)}{2(d+1)\sqrt{z_*^2-z^2}}\right)\,.
\ee
Finally, imposing that $r(0)=R$ we arrive to
\be
R=z_*\left[1+\frac{\varepsilon}{d+1}\left(\frac{z_*}{z_H}\right)^d+\mathcal{O}\left(\frac{z_*}{z_H}\right)^{2(d-1)}\right]\,,
\ee
which can be inverted to obtain
\be\label{zstcorrectedb}
z_*=R\left[1-\frac{\varepsilon}{d+1}\left(\frac{R}{z_H}\right)^d+\mathcal{O}\left(\frac{R}{z_H}\right)^{2(d-1)}\right]\,.
\ee
Unfortunately, if we plug (\ref{zstcorrectedb}) into (\ref{tsatcorrected}) we find that the leading correction to $t_{\text{sat}}$ cancels out, so we have to go even higher order. At the next level, we could not find an analytic solution for $r_{2(d-1)}(z)$.


\end{document}